\documentclass[prb,twocolumn,longbibliography,amsmath,floatfix,superscriptaddress,amssymb,nobibnotes]{revtex4-2}
\usepackage{tikz}
\usepackage{makeidx,amssymb,amsmath,amsthm,graphicx}
\usepackage[toc,page]{appendix}
\usepackage{dcolumn}
\usepackage{bm}
\usepackage{color}
\usepackage{placeins}
\usepackage{epstopdf}
\usepackage{multirow}
\usepackage{rcs}
\usepackage[normalem]{ulem}
\usepackage{chngcntr}
\usepackage[thinspace,thinqspace]{SIunits}
\usepackage{natbib}
\usepackage{hyperref}
\usepackage[hyphenbreaks]{breakurl}

\newcommand{\muB}{$\mu_{\textrm{B}}$}

\newcommand{\YNP}{YbNi$_{4}$P$_{2}$}
\newcommand{\YNA}{YbNi$_{4}$As$_{2}$}
\newcommand{\LNP}{LuNi$_{4}$P$_{2}$}

\newcommand{\YNPA}{YbNi$_4$(P$_{1-x}$As$_{x}$)$_2$}
\newcommand{\YRS}{YbRh$_{2}$Si$_{2}$}
\newcommand{\CRG}{CeRh$_6$Ge$_4$}

\newcommand{\TK}{$T_{\textrm{K}}$}

\newcommand{\TC}{$T_{\textrm{C}}$}
\begin{document}
\title{Physical properties of the ferromagnetic quantum critical system \YNPA}
\author{Kristin~Kliemt}
\email[]{kliemt@physik.uni-frankfurt.de}
\affiliation{Physikalisches Institut, Goethe-Universit\"at Frankfurt/M, 60438 Frankfurt/M, Germany}
\author{Jacintha~Banda}
\affiliation{Max Planck Institute for Chemical Physics of Solids, 01187 Dresden, Germany}
\author{Pavlo~Khanenko}
\affiliation{Max Planck Institute for Chemical Physics of Solids, 01187 Dresden, Germany}
\author{Ali Scherzad}
\affiliation{Physikalisches Institut, Goethe-Universit\"at Frankfurt/M, 60438 Frankfurt/M, Germany}
\author{Ulrike~Stockert}
\affiliation{Technische Universität Dresden, 01062 Dresden, Germany}
\author{Anna~Efimenko}
\affiliation{European Synchrotron Radiation Facility, 38043 Grenoble, France}
\affiliation{Department Interface Design, Helmholtz-Zentrum Berlin für Materialien und Energie GmbH (HZB), Albert-Einstein-Str. 15, 12489 Berlin, Germany}
\author{Kurt~Kummer}
\affiliation{European Synchrotron Radiation Facility, 38043 Grenoble, France}
\author{Cornelius~Krellner}
\affiliation{Physikalisches Institut, Goethe-Universit\"at Frankfurt/M, 60438 Frankfurt/M, Germany}
\author{Manuel~Brando}
\affiliation{Max Planck Institute for Chemical Physics of Solids, 01187 Dresden, Germany}
\date{\today}
%
\begin{abstract}
We report on single crystal growth and physical properties of the quantum critical Kondo-lattice system \YNPA\ with $0\leq x\leq 1$ which hosts a ferromagnetic quantum critical point at $x \approx 0.1$. We performed measurements of the magnetization, electrical resistivity, thermopower, heat capacity, and resonant X-ray emission spectroscopy. Arsenic substitution leads to a homogeneous increase of the unit-cell volume, with well-defined As-concentrations in large parts of the single crystals. All data consistently show that with increasing $x$ the Kondo temperature increases, while the magnetic anisotropy observed at low $x$ fully vanishes towards $x=1$. Consequently, at low temperatures, the system shows a crossover from pronounced non-Fermi liquid behaviour for $x \leq 0.2$ to a Fermi liquid behavior for $x > 0.2$ with weak correlations. There is a continuous change in Yb valence from nearly trivalent at low $x$ to a slightly lower value for $x = 0.6$, which correlates with the Kondo temperature. 
Interestingly, specific heat measurements at very low temperatures show that $C/T$ strongly increases towards lower $T$ for $x = 0.13$ and $x = 0.2$ with a very similar power law. This suggests that in \YNPA\ a quantum critical region rather than a quantum critical point might exist.
\end{abstract}
%
\maketitle
\def\neel{{N\'eel} }
\def\FA{F_{\rm an}}
\def\CA{C_{\rm an}}
\def\MS{M_{\rm sat}}
\def\EDF{E_{\rm df}  }
\def\text#1{{\rm #1}}
\def\i{\item}
\def\[{\begin{eqnarray*}}
\def\]{\end{eqnarray*}}
\def\bv{\begin{verbatim}}
\def\ev{\end{verbatim}}
\def\ganz{Z}
\def\3{\ss}
\def\reel{{\cal}R}
\def\platz{\;\;\;\;}
\def\beginvector{\left(\begin{array}{c}   }
\def\endvector{\end{array}\right)}
\def\fff{\frac{3}{k_B} F }
\def\vec#1{ {\rm \bf #1  } }
\def\KBMUEF{\frac{3k_B}{\mu_{\rm eff}^2 }}
%
\section{Introduction}
Ferromagnetic quantum criticality has been extensively studied in condensed matter research over the last decades (cf. Ref.~\cite{Brando2016} and references therein) and continues to be investigated both experimentally and theoretically. This is evidenced by several recent reports~\cite{Lai2018,Komijani2018,Kotegawa2019,Hafner2019,Shen2020,Kirkpatrick2020,Huang2020,Chen2022,Wendl2022,Jin2022,Shin2023,Thomas2024,Lin2024}.

In metals, general behavior has been identified: The presence of a ferromagnetic (FM) quantum critical point (QCP) is prevented by the coupling of the order parameter to electronic soft modes~\cite{Belitz1999,Chubukov2004,Betouras2005}, resulting in a first-order quantum phase transition, or by the change of the magnetically ordered state into a spin-density-wave/antiferromagnetic (SDW/AFM) state as the QCP is approached~\cite{Brando2008,Conduit2009,Kotegawa2013,Krueger2014,Jesche2017,Hamann2019}. When disorder is large, the quantum phase transition (QPT) is smeared out and spin-glass or quantum Griffiths effects can dominate~\cite{Vojta2010,Westerkamp2009,Lausberg2012,Wang2017}.

To overcome this situation, two relevant proposals have been made: i) In non-centrosymmetric systems, antisymmetric spin-orbit coupling protects the FM QCP~\cite{Kirkpatrick2020}. The FM quantum critical system \CRG\ shows the presence of a FM QCP and is indeed non-centrosymmetric~\cite{Kotegawa2019,Shen2020,Thomas2024}. This concept seems to be also valid for centrosymmetric systems with locally non-centrosymmetric structure, which in combination with non-symmorphic symmetry can protect ferromagnetic QCPs~\cite{Shin2023}. ii) The second possibility is to consider a ferromagnetic metal with quasi-one-dimensional electronic structure in which soft modes are weak. This is the case of the heavy-fermion (HF) system \YNPA\ which shows one-dimensional character of the Fermi surface~\cite{Krellner2011} and the presence of a FM QCP~\cite{Steppke2013}. 

\YNPA\ is the system of the present study. In contrast to Ce-based Kondo-lattice (KL) systems like \CRG\ - in which hydrostatic pressure can be used to suppress ferromagnetism while increasing the Kondo temperature \TK\ - in Yb-based KL systems pressure increases the Curie temperature \TC\ while suppressing \TK. For this reason, in the FM KL compound \YNP\ (with \TC\ = 0.17\,K) arsenic substitution was used as negative chemical pressure to tune this material to its FM QCP at a small As content $x = 0.13$. It was argued that this small amount of As does not introduce enough disorder in the system to suppress quantum critical fluctuations at the QCP and to smear out the QPT. This was confirmed by $\mu$SR experiments showing that magnetism remains homogeneous upon As substitution, without indication of a disorder effect. At the same time, the critical fluctuations are very slow, becoming even slower when approaching the FM QCP~\cite{Sarkar2017}.
However, at that time, it was only possible to grow high-quality single crystals with low As concentrations by a modified Bridgman method which yields thin needle shaped single crystals~\cite{Steppke2013}. Thus, the low-temperature properties of the whole series \YNPA\ and the complete $T-x$ phase diagram are still unexplored. It is, however, important to complete the study. For instance, the highest studied As-concentration $x = 0.13$ presents pronounced non-Fermi-liquid (NFL) behavior at the lowest measured temperature and it is presently unclear how Fermi-liquid (FL) behavior recovers in the series with higher $x$.  Furthermore, in \YNP\ the crystalline electric field (CEF) levels are relatively low in energy~\cite{Huesges2018} and might become of similar order of magnitude as the Kondo energy scale \TK. It is therefore important to check whether the system exhibits a crossover to an intermediate valence state upon As substitution.

Recently, the growth of pure and As-substituted \YNP\ by the Czochralski method was reported~\cite{Kliemt2016,Kliemt2016a}. In contrast to the previous technique, this method allows the growth of larger single-phase crystals. Here, we present the single crystal growth and physical properties of nine single crystals of \YNPA\ with $0 \leq x \leq 1$. 
We show that As substitution leads to a homogeneous increase of the unit-cell volume and a continuous increase of the Kondo temperature, whereas the magnetic anisotropy observed at low $x$ fully vanishes towards $x=1$. Consequently, at low temperatures, the system shows a crossover from pronounced non-Fermi liquid behaviour for $x \leq 0.2$ to a Fermi liquid behavior for $x > 0.2$ with weak correlations as well as a continuous change in Yb valence from nearly trivalent at low $x$ to a slightly lower value for $x = 0.6$. Interestingly, heat capacity data at very low temperatures show that $C/T$ strongly increases towards lower $T$ similarly for $x = 0.13$ and $x = 0.2$, suggesting that in \YNPA\ a quantum critical region rather than a quantum critical point might be present.
%
\subsection*{Previous studies on \YNP}
The tetragonal compound \YNP\ exhibits an extremely low Curie temperature, \TC\ = 0.17\,K~\cite{Krellner2011}, with a strongly reduced ordered moment of $(2.5-4.6) \times 10^{-2}\mu_{\rm B}$, determined by zero field muon-spin relaxation~\cite{Spehling2012}. \TC\ can be suppressed to zero by substituting P with isoelectronic As. The divergence of the Gr\"uneisen ratio at As content $x = 0.08$ was considered evidence of the presence of quantum critical fluctuations at the FM QCP~\cite{Zhu2003,Steppke2013} and drew the attention of the community to this material. This resulted in a series of experimental studies on the pure compound: The CEF level scheme of \YNP\ was determined from neutron scattering, heat capacity, susceptibility and nuclear magnetic resonance data~\cite{Huesges2013,Huesges2018}; The strong ferromagnetic correlations were proven by NMR studies~\cite{Sarkar2012,Baenitz2013}; Multiple Lifshitz transitions at fields between 0.5 and 12\,T were detected~\cite{Pfau2017} and the angular dependence of the critical fields which induce these transitions was studied~\cite{Karbassi2018}. Furthermore, resonant X-ray emission spectroscopy (RXES) on \YNP\ and various Kondo systems was performed~\cite{Kummer2018}, confirming the scaling between the valence at low temperatures and \TK.  Finally, inspired by the surprising finding in \YNP\ ~\cite{Steppke2013} and Yb(Rh$_{0.73}$Co$_{0.27}$)$_{2}$Si$_{2}$~\cite{Lausberg2013,Andrade2014} that, despite a substantial magnetocrystalline anisotropy~\cite{Krellner2012}, the FM moments are oriented along the CEF magnetic hard axis, it was confirmed that almost all KL ferromagnets show this behavior~\cite{Krueger2014,Hafner2019,Rai2019}.
%
\begin{table}[b]
\begin{center}
\begin{tabular}{|c|cccc|}
\hline\hline
$x_{\rm nom} $ &  $\quad x_{\rm EDX} \quad$        &	$\quad a [$\AA$]\quad$       &   $\quad c [$\AA$]\quad$	   	 &  $\quad V [$\AA$^3]\quad$\\
\hline
\hline	
0  & 0	     & 7.0585(1)	   &    3.5888(2)       & 178.80(4)   \\
0.1& 0.13(1)	     & 7.0706(2)	   &	3.6010(7)  	& 180.03(1) \\
0.2& 0.20(1)	     & 7.0838(2)	   &	3.6145(3)  	& 181.37(9)   \\
0.4 &0.40(1)	     & 7.1182(4)	   &	3.6512(1)  	& 185.00(5)  \\
0.6&	0.58(1)     & 7.1500(8)	   &	3.6889(2)  	& 188.59(2)   \\
0.8	&  0.79(1)   & 7.1797(5)	   &	3.7243(1) 	& 191.98(3)  \\
1	&  1   & 7.2035(1)	   &	3.7513(6)  	& 194.66(0)   \\
\hline\hline
\end{tabular}
\end{center}
\caption{\label{Tabelle} Lattice parameters $a$, $c$ and the volume $V$ of the unit cell of \YNPA\ for the nominal As content, $x_{\rm nom}$, (As content from EDX, $x_{\rm EDX}$) determined from X-ray powder diffraction data recorded by using the characteristic line of Mo-radiation of wavelength $\lambda = 0.71069\,$\AA.  } \label{Gitterkonstanten}
\end{table}
%
\section{Experimental Details}
High purity starting materials Yb ingot (99.9\%, Strem Chemicals), Ni slugs (99.995\%, Alfa Aesar), red P pieces (99.999\%, Mining \& Chemical Products Ltd.), black As pieces (99.9999\%, MaTeck) were used. Some of the reagents, namely ytterbium and phosphorous, are air sensitive and arsenic is toxic. The preparation of these reagents was done in a glove box filled with argon. YbNi$_4$(P$_{1-{\it x}}$As$_{\it x}$)$_2$ single crystals with 
$x = 0, 0.1, 0.12, 0.15, 0.2, 0.4, 0.6, 0.8, 1.0$ were grown from a Ni$_{81}$P$_{19}$ flux using a sample to flux ratio of 1:1. For the pure \YNP\, that means that the stoichiometric composition of the elements was weighed in together with 50 at\% Ni$_{81}$P$_{19}$ as flux. For the growth of the single crystals in the substitution series P was partially replaced by As according to the desired composition $x$. The total mass of each growth charge was 15$\,$g. The elements were filled in a boron nitride crucible (V = 30 ml) for the preparation of the precursor. The inner crucible was put in an outer crucible made of niobium which was sealed under Ar using arc-melting. The Czochralski growth experiment was performed in a commercial ADL (Arthur D. Little) high frequency growth device equipped  with a generator that provides a maximum power of about 30 kW. The temperature was measured with an IRCON pyrometer. The crystal structure was characterized by powder x-ray diffraction (PXRD) on crushed single crystals, using Mo-K$_{\alpha}$ radiation. The chemical composition was measured by energy-dispersive x-ray spectroscopy (EDX). For all As substitution levels $x\geq 0.2$, we found a distribution coefficient of $\kappa = c^{As}_l/c^{As}_s=1$ with the concentrations of As in the melt $c^{As}_l$ and in the solid $c^{As}_s$ which was already confirmed for the sample with $x=0.2$~\cite{Kliemt2016a}.
In this manuscript, the given substitution level $x$ refers to the nominal As concentration $x_{\rm nom}$ if not stated otherwise. The orientation of the single crystals was determined using a Laue camera with x-ray radiation from a tungsten anode. 
Four-point resistivity, thermopower, heat capacity and magnetization measurements were performed using the commercial measurement options of a Quantum Design PPMS. The low temperature heat capacity was measured with a semi-adiabatic heat pulse technique~\cite{Wilhelm2004} in an Oxford Instruments $^3$He-$^4$He-dilution refrigerator in a temperature range $0.04$\,K$\, \leq T \leq 5$\,K and magnetic fields up to $2\,$T.
Temperature dependent resonant X-ray scattering (RXES)  measurements at the Yb $L_3$ edge were performed at the ID20 beamline of the ESRF. The experiment was performed in the 16 bunch filling mode of the storage ring, using the Si(111) pre-monochromator of the beamline, five 1\,m radius Si(553) crystal analysers in the spectrometer, and an avalanche photodiode detector \cite{MorettiSala2018}. A continuous He flow cryostat was used for controlling the sample temperature between 6\,K and 300\,K.
%
\section{Crystal growth and characterization}
\subsection{Czochralski growth from a levitating melt}
The crystal growth in the substitution series is performed from a Ni-P, Ni-P-As or Ni-As self-flux as it was described in detail in \cite{Kliemt2016, Kliemt2016a} using the low melting eutectics, Ni$_{80.4}$P$_{19.6}$, with $T_{E,P}=875^{\circ}$C \cite{Huang2010} and Ni$_{76.6}$As$_{23.4}$, with $T_{E,As}=897^{\circ}$C \cite{Massalski} of the binary phase diagrams. The element combination of the high melting transition metal Ni ($T_L$=1455$^{\circ}$C) with non metallic red P (sublimation at 416$^{\circ}$C) and the toxic As (sublimation at 614$^{\circ}$C) as well as of Yb having  a low boiling point (1196$^{\circ}$C) necessitates the preparation of a precursor in a closed Nb crucible. The preparation of the precursor is performed in an inner BN crucible to avoid alloying of the metallic melt with the Nb crucible and comprises three steps: (i) To avoid the contamination of the melt with the crucible material BN, the reaction of P, As with Yb and Ni was performed at comparatively low temperatures. For all compositions, the closed Nb crucible was slowly heated up to 700$^{\circ}$C with a rate of 30 K/h and up to $T_{\rm max}$ with a rate of 50 K/h. The maximum temperatures $T_{\rm max}$ for this process were 950$^\circ$C for the P-rich side and were reduced to 850$^\circ$C for high As contents. This prereaction was done in a box furnace under argon atmosphere.
%
%
(ii) The complete charge was removed from the BN crucible and placed in the cold copper crucible of the high frequency furnace under an Ar pressure of 20 bar. To complete the prereaction and to ensure homogenization the precursor for each composition was heated 15-20 times upon oscillating its temperature between 900$^\circ$C and about $1550^\circ$C. This step was always finished when the reflective surface of the melt was visible and the thin layer of oxides or high melting phases at the surface of the melt slipped down to the bottom of the molten drop-shaped precursor. (iii) In preliminary studies it turned out that the seeding is hindered by a layer consisting of high-melting impurities. We therefore removed the impurity layer mechanically and subsequently by rinsing in ethanol. The three homogenization and cleaning steps caused a weight loss of 1-2\% of the precursor.

The Yb-Ni-P/As melt exhibits a high reactivity with other materials leading to lack of inert crucible material which makes the growth from a levitating melt indispensible. The purified precursor was put in a cold copper crucible (Hukin-type). The Czochralski growth experiment was started by melting the purified precursor with a radio-frequency induction coil applying a power of 14 kW at maximum. After several minutes the precursor was homogenized due to the strong stirring of the levitating melt and the seeding was started. The temperature of the melt before the seeding was chosen to be about $50\,$K above the liquidus temperature. The solidus and liquidus temperatures are summarized in Fig.~\ref{solidliquid}(a) of the Supplemental Material (SM)~\cite{SM}. For the seeding a single crystal seed prepared by a preliminary Czochralski growth experiment with slightly different $x$ was used \cite{Kliemt2016}. 

In all growth experiments, the single crystal seeds were oriented along their crystallographic $[001]$-direction and as soon as the process run stable after dipping, the seed was pulled upwards. 
During the experiments, the total power reduction was about 30\% within the process time of $\approx 50\,$h. 
All samples were grown in the same manner by starting the growth with a pulling rate of $\approx 0.3\,$mm/h and reducing the speed of the pulling rod after 24 h to $0.15\,$mm/h. In a flux-growth process such low growth rates are essential to achieve inclusion-free samples.
After pulling the crystal $\approx 12\,$mm with the low pulling rate, the growth was terminated by pulling faster ($\approx 10\,$mm/h) to separate the sample from the residual melt. Typical growth results are shown in Fig.~\ref{solidliquid}(b).
\subsection{Structural and chemical characterization}
A piece of each \YNPA\ sample was analyzed with PXRD and the measurements confirmed the $P4_2/mnm$ tetragonal structure in the whole substitution series.
The lattice parameters are listed in Tab.~\ref{Tabelle}. The structure refinement using the General Structure Analysis System (GSAS)~\cite{GSAS, EXPGUI} yields an enlargement of the unit cell with increasing As content. The analysis of the PXRD patterns shows that both lattice parameters $a, c$ increase with ${\it x}$ leading to an increase of the unit cell volume V(\YNA) = 1.09 V(\YNP), cf. Fig.~\ref{lattice} in SM. The normalization of the data to the lattice parameters of \YNP\ shows that the $c$-lattice parameter increases stronger than the $a$ parameter. The enlargement of the unit cell for increasing ${\it x}$ is directly visible from the shift of the $(1 0 1)$ reflex to smaller angles as shown in the inset of Fig.~\ref{peakshift}. The increase of the lattice parameters follows Vegards law indicating a stable Yb valence in the series.

The growth of $x=0.1$ was done using an \YNP\ single crystal and for those of $x>0.1$ seeds with slightly lower As contents from the previous growth were used since no respective As substituted seeds with the target substitution were available. It was essential to analyze the chemical composition of the grown single crystals by EDX for two reasons: First, the distribution coefficient of As in the system $\kappa = c_l^{\rm As}/c_s^{\rm As}$ with the concentrations of As in the melt $c_l^{\rm As}$ and in the solid $c_s^{\rm As}$, respectively, is not known and does not necessarily have to be exactly one~\cite{Wilke1988}. This could lead to the enrichment or depletion of As in the melt and an inhomogeneous As distribution in the grown crystal. The analysis was done for each sample  similar to that presented in Ref.~\cite{Kliemt2016a} and showed that the initial As:P ratio of the melt of $x/(1-x)$ can be found all over in our sample except at the first part which is connected to the seed. Second, the As concentration in the seed was always equal or lower than that of the sample. We found that over a length of about 2.5 - 4\,mm the crystal structure included more and more As and the lattice adapted to the new lattice parameters. The analysis along radial lines on polished surfaces, yields a homogeneous As concentration from the center to the surface of the crystal. The As concentration was analyzed with EDX along the $[001]$ direction of all grown crystals to determine the section where the As concentration, $x_{\rm EDX}$, starts to be constant and very close to the nominal concentration, $x_{\rm nom}$, as given in Tab.~\ref{Tabelle}. In the remaining manuscript we, therefore, use $x = x_{\rm nom}$ to indicate the As content.

%
\section{Experimental results and discussion}

\subsection{\YNA}
%
\begin{figure}[ht!]
\centering
\includegraphics[width=\columnwidth]{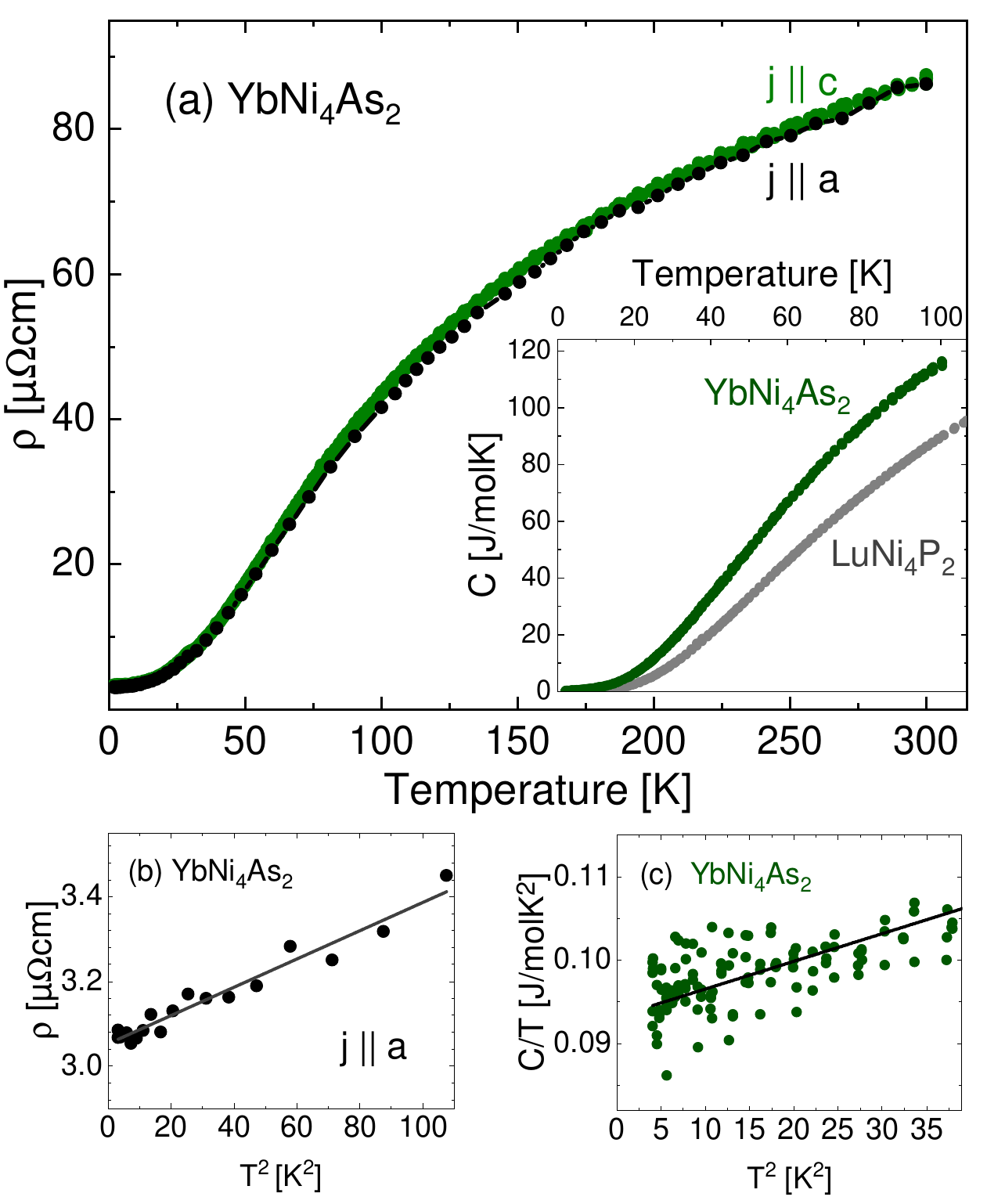}
\caption{(a) Electrical resistivity of \YNA\ measured with currents $j \parallel a$ (black curve) and $j \parallel c$ (green curve). {\it Inset:} Heat capacity of \YNA\ (green) and \LNP\ (gray). 
(b)  Below $10\,\rm K$, for both current directions the resistivity data can be described by $\rho(T)=\rho_0 + AT^2$ indicating FL behaviour. (c)  $C/T$ versus $T^2$ is shown below $6\,\rm K$.}
\label{RhoYNAs}\label{YbNi4As2_HC}
\end{figure}
%
In this section we discuss the properties of \YNA\ single crystals. The temperature dependence of the electrical resistivity $\rho(T)$ is shown in Fig.~\ref{RhoYNAs}(a) of a sample with residual resistivity ratio $RRR_{1.8\,\rm K}=27$. $\rho(T)$ decreases monotonically with temperature showing a slight curvature with no sign of a single-ion Kondo effect and is isotropic for both current directions ($j \parallel a, j \parallel c$). This is in contrast to \YNP\ which shows strongly anisotropic values: $RRR_{1.8\,\rm K}=17$ ($j\parallel c$) and $RRR_{1.8\,\rm K}\approx 3$ ($j\perp c$) \cite{Kliemt2016}. Below $10\,\rm K$, $\rho(T)$ of \YNA\, shows Fermi liquid behaviour as $\rho(T)$ is well described by $\rho(T)=\rho_0 + AT^2$ with  $A = 0.0033\,\mu\Omega\rm cm/K^2$ and $\rho_0=3.05\,\mu\Omega\rm cm$. This is shown in Fig.~\ref{RhoYNAs}(b).

The heat capacity $C(T)$ of \YNA\ is presented in the inset of Fig.~\ref{YbNi4As2_HC}(a) for temperatures between 2 and 100\,K, together with the heat capacity of \LNP~\cite{Kliemt2018}. There is a large difference in $C(T)$ at high temperature: In a simple approximation the Debye temperature $\Theta_D \propto 1/\sqrt{m}$, where $m$ is the average atomic mass ~\cite{Anderson1965}. So, we do expect a rather small change of $\Theta_D$ upon P/As exchange since the mass is dominated by Ni and Yb/Lu. But this is not the case here and therefore a substantial error occurs due to the difference in the atomic mass when using the \LNP\ data as a non-magnetic phonon reference for \YNA. Below 6\,K, the heat capacity of \YNA\ shows FL behavior since it can be described by $C/T = \gamma_0 + \beta T^2$ (see Fig.~\ref{YbNi4As2_HC}(c)) with a moderately enhanced Sommerfeld coefficient $\gamma_0 = 96\,\rm mJmol^{-1}K^{-2}$. A Debye temperature $\Theta_D = (346\pm 5)$\,K was determined from the slope $\beta=12\pi^4Rs/5\Theta_D^3$ (with the gas constant $R$ and $s=7$). The latter value is considerably smaller than the one determined for \LNP, $\Theta_D = (420\pm1)$\,K~\cite{Kliemt2018}.

\begin{figure}[b]
\includegraphics[width=1.0\columnwidth]{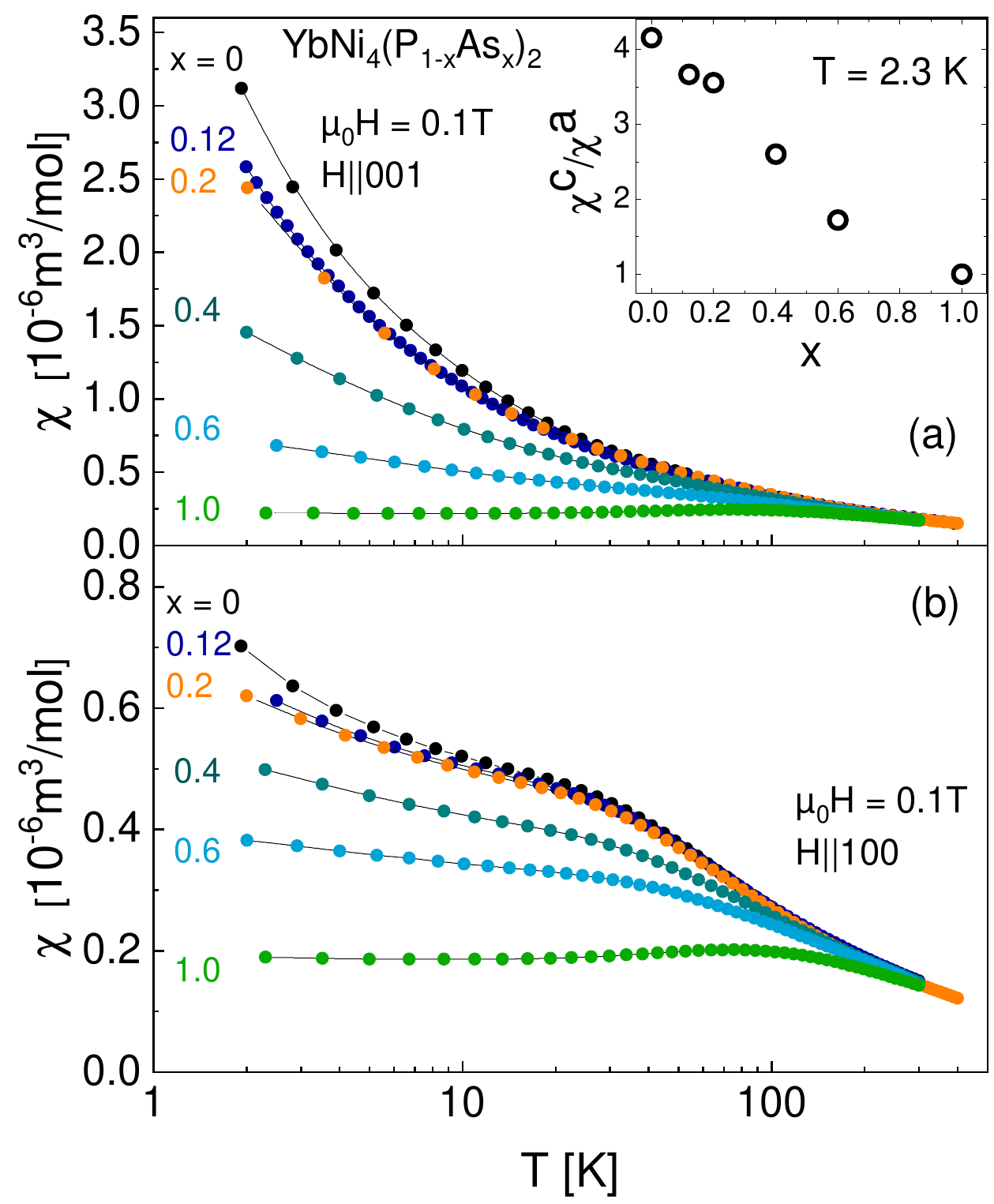}
\caption{\YNPA: Temperature dependence of the molar magnetic susceptibility $\chi(T)$ for $\mu_{0}H=0.1\,\rm T$ with (a) $H\parallel 001$ and (b) $H\parallel 100$. The inset shows the low-$T$ anisotropy of $\chi(T)$ for the different $x_{\rm nom}$ between the two field directions at $\mu_0H = 0.1$\,T determined from the data at $T = 2.3$\,K. 
}
\label{MvT_YBNi4PAs2}
\end{figure}

A first characterization of polycrystalline \YNA\ was reported by Deputier {\it et al.}~\cite{Deputier1997} showing a strong increase of the susceptibility towards low temperatures. We cannot reproduce this behaviour and speculate that this strong increase could be caused by a secondary phase contribution. The absolute values of the susceptibility of the As compound are much smaller than those of the P compound (cf. Fig.~\ref{MvT_YBNi4PAs2}). The systematic behavior of the susceptibility in \YNPA\ indicates a pronounced increase of the Kondo interaction together with a reduction of the FM correlations with increasing $x$ (cf. section below). A Curie-Weiss (CW) behaviour above 200\,K is observed with Weiss temperatures $\Theta_W^c= (-115\pm 5)\,\rm K$ and $\Theta_W^a= (-149\pm 5)\,\rm K$ indicating prevalent antiferromagnetic exchange (see Fig.~\ref{MvT_YBNi4PAs2_1a}(b) of the SM). The field dependent magnetization of \YNA\ taken at 2\,K with the magnetic field parallel and perpendicular to the $c$-axis is isotropic (see inset of Fig.~\ref{MvT_YBNi4PAs2_1a}(b) of the SM) in contrast to the strong anisotropy observed in \YNP~\cite{Steppke2013,Gegenwart2015}.
Summarizing this basic characterization for $x=1$, \YNA\, can be classified as an intermetallic Kondo system with isotropic electrical transport properties.  
%
%
\subsection{YbNi$_4$(P$_{1-{\it x}}$As$_{\it x}$)$_2$}

\subsubsection{Magnetic susceptibility}
In this section, we discuss the magnetic properties of the \YNPA\ single crystals. The magnetic susceptibility $\chi (T)$ of all crystals is shown in Fig.~\ref{MvT_YBNi4PAs2} for temperatures between 2 and 400\,K. Above 200\,K all curves show CW behavior for both field directions ($H \parallel 001$ and $H \parallel 100$) with considerable magnetic anisotropy for the systems with low $x_{\rm nom}$ at low temperatures and effective moments close to 4.54\,\muB\, which corresponds to the $J = 7/2$ multiplet of the free Yb$^{3+}$ ion. This indicates that Ni is non magnetic for all compounds. It has to be mentioned that for high As concentrations, a constant Pauli susceptibility $\chi_{0}$ had to be included in the CW fits. This is also justified by the XAS and RXES data (cf. Sec.~\ref{spectroscopy}) 
which show that the Yb valence for all samples up to $x = 0.6$ is higher than 2.975+ for $T > 200$\,K.
%
\begin{table}
\begin{center}
\begin{tabular}{|c|cccc|}
\hline\hline
$x_{\rm nom} $ &  $\quad \Theta_W^{a} \quad$        &	$\quad \Theta_W^{c}\quad$       &   $\quad B^0_2\quad$	   	 &  $\quad $Ref.$\quad$\\
&$\pm 5 [\rm K]$&$\pm 5 [\rm K]$&$\pm 0.05 [\rm meV]$&\\
\hline
\hline	
0   &-33&-7&  -0.13 & \cite{Krellner2012}\\
0.12&-46&-16&	-0.14  	& this work \\
0.2 &-48&-16&	-0.15  	& this work \\
0.4 &-62&-25&	-0.18 	& this work \\
0.6 &-72&-43&	-0.14  	& this work \\
1	&-149&-115&	-0.16  	& this work \\
\hline\hline
\end{tabular}
\end{center}
\caption{\label{TabelleThetaW} Weiss temperatures of YbNi$_4$(P$_{1-{\it x}}$As$_{\it x}$)$_2$ determined from the inverse susceptibility $\chi^{-1}(T)$ and the first CEF parameter $B^0_2$ calculated according to Eqn.~(\ref{B02}). See also Fig.~\ref{MvT_YBNi4PAs2_1a}c in the SM. }
\label{Weisstemperatures}
\end{table}
%
From the high-temperature CW fits we have extracted the Weiss temperatures $\Theta_W^{a}$ and $\Theta_W^c$. The values are listed in Tab.~\ref{TabelleThetaW}. They are all negative indicating predominant AFM intersite exchange interactions. Although the absolute values systematically increase with increasing $x$, indicating an enhancement of the AFM exchange, their difference changes only slightly (cf. Fig.~\ref{MvT_YBNi4PAs2_1a}c in the SM), indicating almost a constant value of the leading CEF parameter $B_{2}^{0}$. 
In fact, based on molecular field theory~\cite{Bowden1971,Boutron1973,Klingner2011}, the first CEF parameter $B^0_2$ - being a measure of the strength of the magnetocrystalline anisotropy - can be expressed according to
%
\begin{equation}\label{B02}
    B^0_2=(\Theta_W^{a}-  \Theta_W^c)\frac{10k_{\rm B}}{3(2J-1)(2J+3)}.
\end{equation}
%
This can be explained by the fact that in systems with a stable Yb$^{3+}$ state, due to a small de Gennes factor, magnetic intersite exchanges are much weaker than CEF effects. Therefore, the fit to the high-temperature data reflects the CEF strength, whereas the low-temperature susceptibility reflects the sum of all exchange interactions. Below 50\,K, the susceptibilities for samples with low As content become very high and strongly anisotropic. On the other hand, they remain small and nearly isotropic for samples with higher As concentrations. This is shown in Fig.~\ref{MvT_YBNi4PAs2} and in the inset in which we plotted the rate $\chi^{c}/\chi^{a}$ between the susceptibility values at $T = 2.3$\,K. This anisotropy rate amounts to $\chi^{c}/\chi^{a} = 4.15$ for $x = 0$ while it is $\chi^{c}/\chi^{a} = 1.01$ at $x = 1$. The strong enhancement of the low-$T$ susceptibilities reflects the enhancement of FM exchange interaction with decreasing $x$.


The behavior of the susceptibility is also reflected in the field dependent magnetization $M(H)$ which is plotted for $H \parallel [001]$ and $[100]$ at $2\,\rm K$ in Fig.~\ref{MvT_YBNi4PAs2_1} and Fig.~\ref{MvT_YBNi4PAs2_1a} of the SM. The anisotropy in $M(H)$ decreases with increasing $x$ leading to a nearly isotropic field dependence for $x = 1.0$. For \YNP\ we observe an identical field dependence as it was determined for Bridgman grown single crystals~\cite{Krellner2012}: For $H \parallel [001]$ a polarized moment of 1.24\,\muB\ is measured at $\mu_0H = 7$\,T while for $H \parallel [100]$ the value is only 0.53\,\muB\ at the same field. Most importantly, the polarized moment in both field directions decreases strongly with increasing $x$ evidencing the increasing Kondo screening towards low temperatures. In fact, in \YNA\ the polarized moment at 7\,T is only 0.1\,\muB. 
%
%
\subsubsection{Electrical resistivity}

\begin{figure}
\centering
\includegraphics[width=0.97\columnwidth]{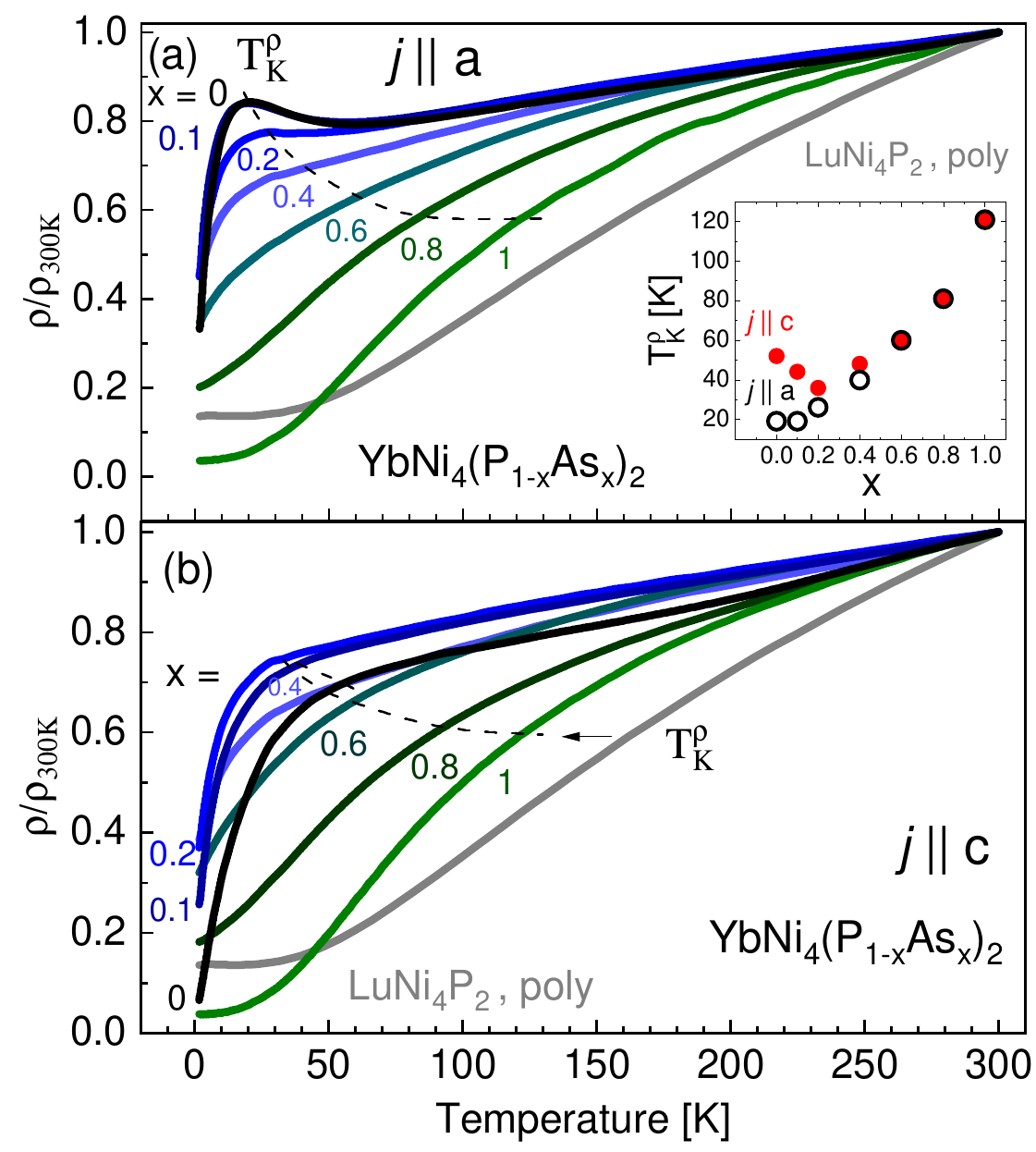}
\caption{Temperature dependence of the normalized electrical resistivity $\rho(T)/\rho_{300\,\rm K}$ for samples with different $x_{\rm nom}$ from 1.8 to 300\,K measured (a) with current $j$ parallel to the crystallographic $a$-axis, $j \parallel [100]$, and (b) parallel to the $c$-axis, $j \parallel [001]$. The grey curve shows the electrical resistivity of the non-magnetic reference \LNP~\cite{Kliemt2016a}. The dashed lines indicate the position of the maximum which is observed in the $\rho_{\rm mag}(T)$ data after subtraction of the non-magnetic reference (cf. Fig.~\ref{RhoARhoC_1}), and attributed to the Kondo temperature $T_{K}^{\rho}$. The shift of $T_{K}^{\rho}$ with $x_{\rm nom}$ for both current directions is plotted in the inset. 
}
\label{RhoARhoC}
\end{figure}

In this section we present measurements of the $T$-dependent electrical resistivity $\rho(T)$ of all \YNPA\ single crystals taken between 1.8\,K and 300\,K and with current $j$ parallel to the crystallographic $a$-axis, $j \parallel [100]$, and parallel to the $c$-axis, $j \parallel [001]$. In Figs.~\ref{RhoARhoC}(a) and (b) the electrical resistivity $\rho(T)/\rho_{300\,\rm K}$ normalized by its value at 300\,K is plotted, for comparison. The grey curve is the resistivity $\rho^{\rm Lu}(T)$ measured on a polycrystalline sample of \LNP\ that we take as non-magnetic reference compound to estimate the phonon contribution to the resistivity~\cite{Kliemt2016a}. 

For $j \parallel [100]$, a maximum is clearly visible in the resistivity data for $x\leq 0.2$ which broadens for higher arsenic concentrations. This is much better visible in a plot of 
%
\begin{eqnarray}\label{rholabel}
\rho_{\rm mag}=\frac{\rho({\rm YbNi}_4{\rm P}_2)}{\rho({\rm YbNi}_4{\rm P}_2, 300\,\rm K)}-\frac{\rho({\rm LuNi}_4{\rm P}_2)}{\rho({\rm LuNi}_4{\rm P}_2, 300\,\rm K)}
\end{eqnarray}
which we use to estimate the magnetic contribution to the resistivity.

%
In $\rho_{\rm mag}(T)$, the maximum can be traced for all arsenic concentrations and with current along both crystallographic directions (cf. Fig.~\ref{RhoARhoC_1} of the SM). We give here no absolute values for $\rho_{\rm mag}(T)$ since the polycrystalline non-magnetic reference has a lower $RRR$ than the single crystalline samples and the difference becomes negative.

The appearance of a maximum resistivity in KL systems is caused by the interplay between the CEF energy level splitting and the onset of the coherence effect related to the Kondo temperature \TK. This well known behaviour has been thoroughly studied, e.g., in CeRu$_2$Ge$_2$~\cite{Wilhelm2005} and CeCu$_2$Si$_2$~\cite{Franz1979} under pressure or in substitution series of \YRS~\cite{Koehler2008,Klingner2011a}. A theoretical description of the resistivity of Ce-based compounds with and without magnetic order at low temperatures including CEF and Kondo effect can be found in Ref.~\cite{Lassailly1985}.
%
%
\begin{table*}
\begin{tabular}{|c|c|c|c|c|c|c|c|c|c|}
\hline\hline
$\quad x   \quad$       &$A^{j\parallel a}$&$\quad T^{\rho,a}_{\rm FL}\quad$&$A^{j\parallel c}$&$\quad T^{\rho,c}_{\rm FL}\quad$         & $\gamma_0$      & $\quad T^{\rm HC}_{\rm FL}\quad$ &$A^{j\parallel a}/\gamma_0^2 $ &$A^{j\parallel c}/\gamma_0^2 $&$T_{\rm K}$\\[+0.2em]
	  &$[\frac{\mu\Omega\rm cm}{\rm K^2}]$&$[\rm K]$&$[\frac{\mu\Omega\rm cm}{\rm K^2}]$&$[\rm K]$&  $[\frac{\rm J}{molK^2}]$&$[\rm K]$&$[\frac{\mu\Omega\rm cmK^2mol^2}{J^2}]$&$[\frac{\mu\Omega\rm cmK^2mol^2}{J^2}]$&$[\rm K]$\\[+0.1em]
\hline\hline
0.2       & $-$ &$-$      & $-$ &$-$&$>2.8$&$-$&$-$&$-$&9.5\\[+0.1em]
0.4       & $-$  &$-$       & $-$  &$-$&0.99&$0.06$&$-$&$-$&18\\[+0.1em]
0.6       &0.59  & $3$&0.19 &3         &$\approx 0.4$&$\approx 0.35$&$\approx 3.7$&$\approx 1.2$&26.5\\[+0.1em]
0.8       &0.07  &$10$&0.037 &6                  & 0.16       & $5$&2.7&1.4&35.4\\[+0.1em]
1         &0.0033&$10$&0.0032&10         &0.096                         &$6$&0.36&0.35&43.8\\[+0.1em]
\hline  \hline
\end{tabular}
\caption{Summary of characteristic parameters of \YNPA : Resistivity $A$ coefficient determined for \YNPA\ with $x\geq 0.6$. $T^{\rho}_{\rm FL}$ is the temperature below which the resistivity can be described by $\rho(T)=\rho_0+AT^2$.  
$T^{\rm HC}_{\rm FL}$ is the temperature below which $C/T$ saturates and follows $C/T(T) = \gamma_0+\beta T^2$ with the Sommerfeld coefficient $\gamma_0$. The Kadowaki-Woods ratio $A/\gamma_0^2 $ was evaluated for both current directions. $T_{\rm K}$ is the Kondo temperature determined from entropy. }
\label{Resistivitycoeff}\label{Sommerfeld}
\end{table*}

In Yb-based Kondo lattice systems, the Kondo temperature decreases with increasing pressure~\cite{Bauer1995} while the CEF levels are typically not affected by pressure significantly~\cite{Koehler2008,Klingner2011a}. For instance, in \YRS\ under pressure besides the existence of a Kondo maximum due to scattering on the ground state also the occurrence of a high-temperature maximum in the electrical resistivity was observed and attributed to inelastic Kondo scattering on excited CEF levels~\cite{Dionicio2005}. Our susceptibility measurements also show that in \YNPA\ the leading CEF parameter does not change drastically with increasing $x$ whereas the position of the maxima for both current directions shifts substantially within the substitution series. This is indicated by dashed lines in Fig.~\ref{RhoARhoC}. So, we attribute this maximum to the coherence Kondo temperature $T_{\rm K}^{\rho}$ because it is unlikely due to a drastic change of the CEF energy level splitting with $x$.

When looking carefully at the data for the pure P compound in~Fig.~\ref{RhoARhoC_1}, a weak shoulder in $\rho_{\rm mag}(T)$ can be seen at high temperatures, around 80\,K for both current directions and with increasing As  content this shoulder broadens and seems to be superposed by the maximum at $T_{\rm K}^{\rho}$. This leads to the conclusion that the Kondo interaction and CEF effects are responsible for the detailed temperature dependence in the electrical resistivity, but it is not possible to clearly separate both effects. This is also evidenced by the thermopower measurements shown in the next section.

Interestingly, the behavior of $T_{\rm K}^{\rho}$ at low arsenic concentrations is anisotropic: The inset of Fig.~\ref{RhoARhoC}(a) shows a strong anisotropy in the position of $T_{\rm K}^{\rho}$ with respect to the two current directions. This anisotropy disappears at higher arsenic concentrations. More specifically, we observe that for $j \parallel c$, the Kondo maximum shifts towards lower temperatures when approaching the quantum-critical arsenic concentration near $x\approx 0.2$. This behavior seems to be not associated with the magnetic anisotropy, because it decreases linearly with increasing $x$ (cf. inset of Fig.~\ref{MvT_YBNi4PAs2}). Instead, it might be interpreted as caused by anisotropic Kondo hybridization, as found in \CRG~\cite{Wu2021}, but we do not have evidence for this, yet. Moreover, part of this feature could be due to the strong difference in the residual resistivity ratio $RRR_{1.8\,\rm K} = \rho(300\,\rm K)/\rho(1.8\,\rm K)$ and the associated drop in $\rho(T)$ at low temperatures. The RRR is regarded as an indicator for the amount of crystal defects and disorder. In the \YNPA, As (P) atoms act as impurities in the P-rich (As-rich) single crystals and cause disorder. The disorder reaches its maximum for $x = 0.4$ where $RRR_{1.8\,\rm K}$ is minimal, as shown in Fig.~\ref{RhoARhoC_1}(a) and (b).

At very low temperatures, Fermi liquid behaviour is observable in \YNPA\ for $x\geq\,0.6$. Here, the resistivity can be described by $\rho(T)=\rho_0+AT^2$. The coefficients extracted from the fits are listed in Tab.~\ref{Resistivitycoeff}. The $A$ coefficient decreases in the substitution series for increasing arsenic content. Since $A\propto (m^*)^2$, we find, as expected, that correlation effects are stronger for lower arsenic contents. 
\subsubsection{Thermopower}
%
\begin{figure}[b]
\includegraphics[width=1.0\columnwidth]{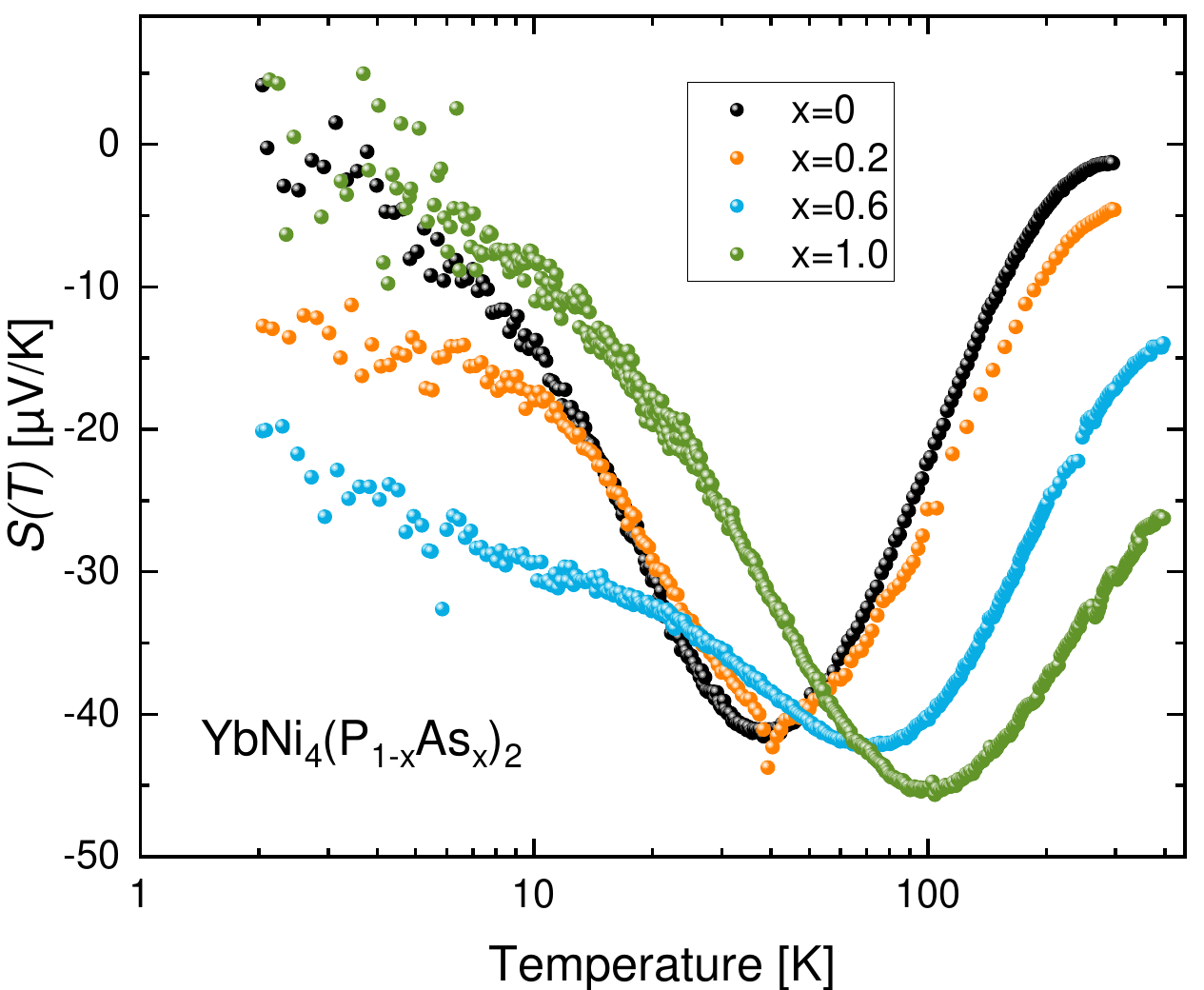}
\caption{Temperature dependence of the Seebeck coefficient $S(T)$ for four crystals of \YNPA.}
\label{Seebeck}
\end{figure}
The temperature dependence of the Seebeck coefficient $\mathcal{S}(T)$ of the pure compound \YNP\ was investigated in Ref.~\cite{Krellner2011} and showed negative values within the temperature range between 2 to 300\,K. This is well established in Yb-based Kondo lattices that order magnetically ("type a" systems in Ref.~\cite{Zlati2005}). In particular, it was found that $\mathcal{S}(T)$ of \YNP\ shows a pronounced minimum at 35\,K with $\mathcal{S}(35\,\rm K)=-40 \mu \rm VK^{-1}$~\cite{Krellner2011}. Such large absolute values near a minimum are usually ascribed to a strongly energy-dependent quasi-particle density of states at the Fermi level~\cite{Zlati2005}. However, the maximum predicted for "type a" systems at low temperature could not be resolved in our measurements down to 2\,K, but it might occur at lower temperatures. The minima in $\mathcal{S}(T)$ originate from Kondo scattering on the ground and excited CEF levels like the maxima observed in $\rho(T)$. This was investigated in detail, for instance, by applying negative chemical pressure through Lu- or Co-substitution on \YRS~\cite{Koehler2008,Klingner2011a,Stockert2019}.

In Fig.~\ref{Seebeck}, we show the measured $\mathcal{S}(T)$ of four single crystals of \YNPA\ measured in the temperature range between 2 and 400\,K. We observe a single negative minimum for all investigated concentrations with only slightly increasing absolute values for higher As content up to $\mathcal{S}_{\rm min}=-45 \mu \rm VK^{-1}$ for \YNA. More importantly, the temperature of the minimum $T_{\rm min}$ changes from $38\,\rm K$ for $x=0$ (which is comparable to 35\,K determined on previous samples~\cite{Krellner2011}), $39\,\rm K$ for $x=0.2$, $72\,\rm K$ for $x=0.6$ up to $104\,\rm K$ for \YNA. This strong shift is related to the shift of the Kondo temperature to higher value on applying negative chemical pressure. The contribution from the CEF levels can not be separated from that due to the Kondo effect on the ground state doublet. 
%
%
\subsubsection{Heat capacity}
Previous work on the series \YNPA\ revealed the existence of a ferromagnetic QCP for $x\approx 0.1$. The heat capacity shows a divergence of $C(T)/T\propto T^{-0.43}$ in samples up to a substitution level of $x = 0.13$ which was the highest As concentration investigated so far~\cite{Steppke2013}. From the heat capacity of our samples with $x \geq 0.15$ we can now obtain further information on the phase diagram, like the onset of FL behaviour and the evolution of the Kondo temperature for higher $x$. 
Here, we present measurements of the $T$-dependent heat capacity $C(T)$ of \YNPA\ taken from 100\,K down to 30\,mK. From all data we have subtracted the nuclear contribution $C_{n} = \alpha_{n}T^{-3}$ due to the presence of the two isotopes $^{171}$Yb and $^{173}$Yb with nuclear spin $I =1/2$ (abundance 14.3\%) and $I = 5/2$ (abundance 16.1\%), respectively~\cite{Steppke2010,Steppke2013}. Due to the strong NFL behavior observed at the lowest temperatures for $x \leq 0.2$ it is difficult to extract this contribution directly from a fit to the data at zero field. However, in magnetic field FL behavior is recovered and a reliable coefficient $\alpha_{n}$ can be estimated from the extrapolation of the high-field data to zero field, since the nuclear energy splitting is proportional to the magnetization ($\alpha_{n} \propto M^{2}$) due to the strong hyperfine coupling~\cite{Steppke2010}. The procedure is described in the SM of Ref.~\cite{Steppke2013} and the data in magnetic field are shown in Figs.~\ref{HC20proz_nucl} and \ref{HC40proz_nucl}.  

\begin{figure}
\centering
\includegraphics[width=1.0\columnwidth]{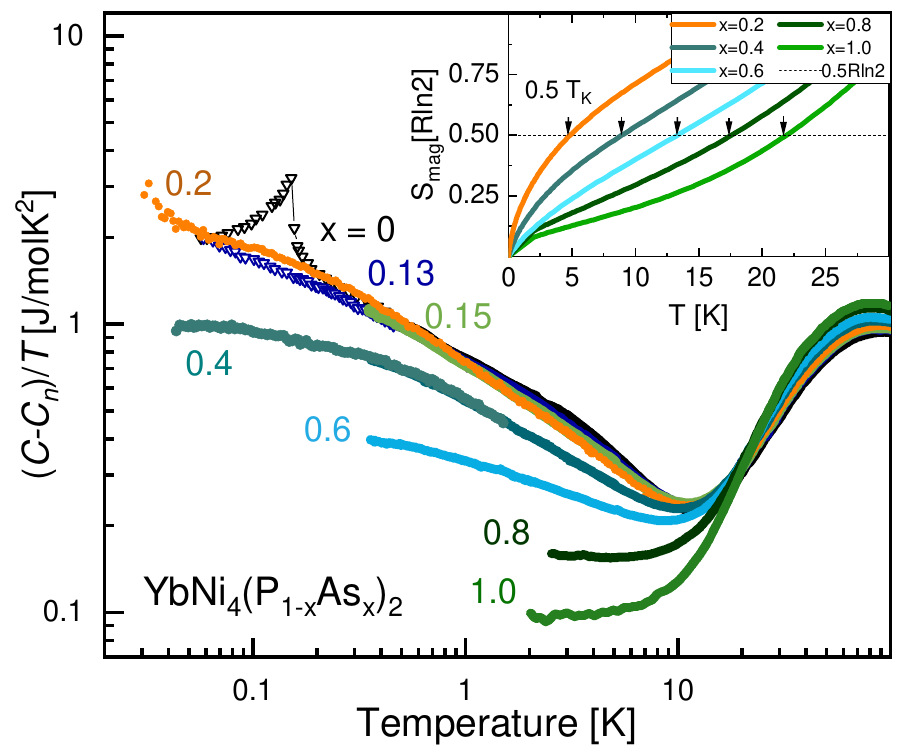}
\caption{$(C-C_n)(T)/T$ of \YNPA\ for $0\,\leq\,x\,\leq\,1$ from 35\,mK to 100\,K (closed circles). Low temperature data (from Ref.~\cite{Steppke2013}) for $x = 0$ (black open triangles) and $x = 0.13$ (blue open triangles) are also plotted for comparison. The inset shows the magnetic part of the entropy $S_{\rm mag}$ which is used to determine the thermodynamic Kondo temperature \TK. The dashed line marks the value of $0.5\,R\rm ln2$, arrows indicate the temperatures where $T=0.5\,T_K$.} 
\label{HCYNPAs}
\end{figure}

The heat capacity divided by temperature $C/T$ is plotted in Fig.~\ref{HCYNPAs}. For $x$ = 0.4, 0.6, 0.8 and 1, $C/T$ flattens towards low temperatures and FL behavior is observed below $0.06\,\rm K$, $\approx 0.35\,\rm K$, $5\,\rm K$ and $6\,\rm K$, respectively. The corresponding enhanced Sommerfeld coefficients are $\gamma_0=992\,\rm mJ/molK^2$, $\approx 400\,\rm mJ/molK^2$, $158\,\rm mJ/molK^2$ and $96\,\rm mJ/molK^2$. $\gamma_{0}(x)$ decreases dramatically from being higher than 2\,J/molK$^{2}$ for $x=0.2$ to 0.1\,J/molK$^{2}$ for $x=1$ as shown in Tab.~\ref{Sommerfeld}. 

For $x \leq 0.2$, $C/T$ increases below 10\,K with a similar power-law $\propto T^{-0.43}$ for all samples. In particular, for $x = 0.2$, $C/T$ reaches values close to 2.8\,J/molK$^2$ and no saturation is observed down to the lowest point at about 35\,mK. Previous investigations have shown a linear decrease of the Curie temperature to zero upon substitution with As at $x \approx 0.1$ associated with NFL behavior in $C(T)/T$ and also in the Gr\"uneisen ratio~\cite{Steppke2013}. For $x > 0.1$ FL behavior is expected. However, neither the crystal with $x = 0.13$ nor the one with $x = 0.2$ show FL behavior. Through EDX analysis \cite{Kliemt2016a}, we determined $x_{\rm EDX}=0.20(1)$ for the sample still showing a divergence which is clearly distinguishable from a $x_{\rm EDX}=0.13(1)$ sample. Also, within one grown crystal, the As concentration is homogeneous \cite{Kliemt2016a}. 

This wide range of As concentrations for which divergence is observed hints to the presence of a quantum critical region rather than a quantum critical point in \YNPA. In disordered systems near a FM QCP it is possible to observe quantum Griffith (QG) effects which are characterized by a power-law increase of the magnetic susceptibility and specific heat coefficient towards low temperatures: $\chi'(T) \propto C(T)/T \propto T^{\lambda-1}$, where $\lambda$ ($0 \leq \lambda \leq 1$) strongly changes with $x$~\cite{Vojta2010,Westerkamp2009,Lai2018}. In \YNPA, however, the parameter $\lambda \approx 0.57$ is almost the same for all samples investigated ($x = 0$, 0.04, 0.08, 0.13 and 0.2), even for the pure \YNP~\cite{Steppke2013}. This clearly rules out strong disorder effects and the presence of a QG phase in our system.  
%

Now, we discuss the contribution of the $4f$ electrons to the entropy $S_{\rm mag}$ which was determined by integration of $C_{\rm mag}/T$ over $T$ from the lowest measured temperature. This is shown in the inset of Fig.~\ref{HCYNPAs}. The Kondo temperatures \TK\ were determined from $S_{\rm mag}$, as \TK$= 2\cdot T(S=0.5 R\,\rm ln 2)$~\cite{Pikul2012} and are listed in Tab.~\ref{Resistivitycoeff}. Note that for high As concentrations  $C_{\rm mag}$ is overestimated due to the subtraction of the heat capacity of \LNP\ which was used as a phonon reference (see for comparison Fig.~\ref{YbNi4As2_HC}). This leads to a slightly underestimated \TK\ for high As concentrations. Nevertheless, as a clear trend we observe an increase of \TK\ with increasing As content, as expected. Furthermore, since the entropy $S_{\rm mag}(T)$ does not exceed $R\,\rm ln 2$ at $10\,\rm K$ for $0.2\leq x \leq 1.0$, we can conclude that only the lowest CEF doublet is involved in the formation of the ground state while the first excited state remains energetically well separated. The same was deduced for the pure phosphorous compound \YNP\ through heat capacity measurements~\cite{Krellner2012} and confirmed in inelastic neutron scattering data~\cite{Huesges2013,Huesges2018}.
%
\subsubsection{Spectroscopy \label{spectroscopy}}

\begin{figure}[t]
\centering
\includegraphics[width=1.0\columnwidth]{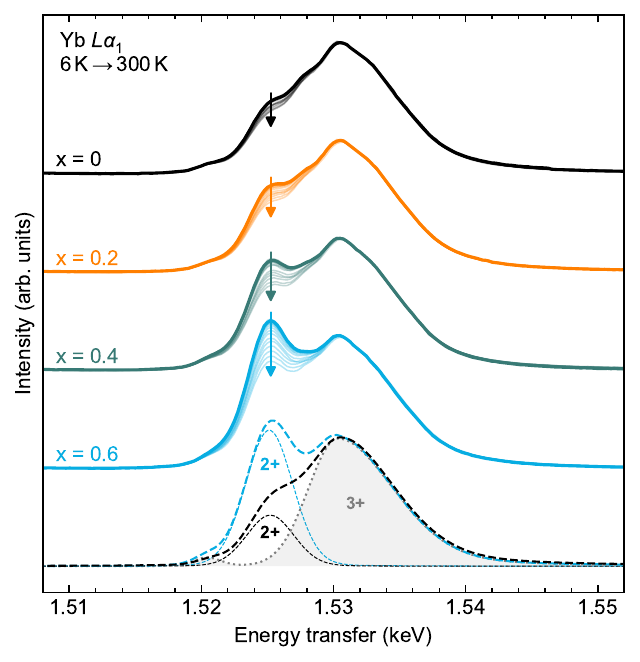}
\caption{\YNPA\,: Temperature evolution of the Yb $L_3$-$L\alpha_1$ RXES spectra. The incident energy was set to $E_\mathrm{in} = 8.9405\,$keV, the maximum of the Yb$^{2+}$ $L_3$ absorption peak, in order to maximize the Yb$^{2+}$ contribution to the spectra. A clear Yb$^{2+}$ peak is seen for all doping levels at low $T$ (thick lines), which increases with As doping and decreases with temperature. All spectra were normalised to Yb$^{3+}$ peak intensity. The relative weight of the Yb$^{2+}$ and Yb$^{3+}$ contributions was used to determine the Yb valence as a function of doping and temperature as described in \cite{Kummer2011, Kummer2018}. The decomposition of the low temperature spectra for $x=0$ and $x=0.6$ is shown as an example at the bottom.}
\label{RIXS1}
\end{figure}

Due to its almost full $4f$ shell (one hole), Yb in compounds may exhibit different valence states depending on the respective chemical composition or other parameters like temperature or pressure~\cite{Lawrence1981}. Its valence instability is connected to the hybridization between localized $4f$ electrons and the conduction electrons~\cite{Brandt1984}. While weak hybridization leads to a stable 3+ configuration and magnetic order of Yb atoms, increasing hybridization results in a screening of the local moments by the conduction electrons in Kondo lattice or HF 
systems~\cite{Doniach1977}. It is known that in such systems the occupation of the $4f$ shell, namely the number of $4f$ holes $n_h$, decreases weakly with increasing hybridization and the valence $v$ slightly deviates from 3+. Recently, it was found that the amount of the deviation scales with the Kondo temperature~\cite{Kummer2018}.
%
\begin{figure}[t]
\centering
\includegraphics[width=1.0\columnwidth]{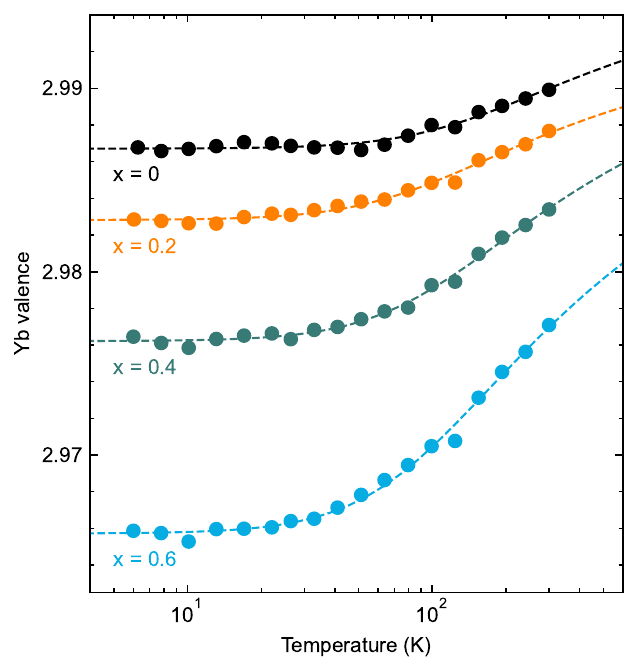}
\caption{\YNPA\,: Change of the Yb valence with temperature as determined from the RXES spectra in Fig.~\ref{RIXS1}.}
\label{RIXS2}
\end{figure}
The valence of Yb ions can be determined using core-level spectroscopies. RXES at the Yb $L_3$ edge is particularly suited in this case because it can detect even small deviations from the Yb$^{3+}$ configuration with very high sensitivity~\cite{Dallera2002,Kummer2011,Kummer2018}. In Fig.~\ref{RIXS1} we show the Yb $L_3$-$L\alpha_1$ RXES spectra for pure \YNP\ and three different As substitution levels $x = 0.2, 0.4, 0.6$ at $T = 6\,K$ and their temperature dependence up to 300\,K. All spectra were normalized to the Yb$^{3+}$ peak intensity. 
The Yb valence can be extracted from the relative weight of Yb$^{2+}$ and Yb$^{3+}$ contributions to the spectra as described previously in Refs.~\cite{Kummer2011,Kummer2018}. The results are shown in Fig.~\ref{RIXS2}. The temperature scale of the valence change $T_v$ is very similar for all samples despite the very different Kondo temperatures \TK. In contrast, the deviation of the low temperature Yb valence from 3+ was found to be well correlated with \TK. This is also confirmed for \YNPA. It can be nicely seen that the deviation from the trivalent state $n_h(0)$ becomes larger with increasing substitution level and increasing Kondo temperature.
%
\subsubsection{Kadowaki-Woods ratio and characteristic energy scales}
In Fermi liquids, the temperature dependence of the resistivity at low temperatures  can be described by $\rho(T)=\rho_0+AT^2$, where $\rho_0$ describes the impurity scattering. The quadratic temperature dependence arises due to electron-electron scattering with a constant $A$ which is proportional to the square of the effective mass $m^*$. The low temperature heat capacity of Fermi liquids can be described by $C(T)=\gamma_0 T$ with the Sommerfeld coefficient $\gamma_0\propto m^*$. The ratio $A/\gamma_0^2$ is known as Kadowaki-Woods (KW) ratio and it was found to take the universal value of $10\,\mu\Omega\rm cmK^2mol^2/J^2$ for HF compounds~\cite{Kadowaki1986}. More recent works have shown that this ratio can be better calculated if the level of degeneracy of quasiparticles $N$ is considered. In HF systems this number depends on the interplay between \TK\ and the CEF energy splitting $\Delta_{CEF}$. If \TK\ $< \Delta_{CEF}$, the low-temperature properties can be explained by $N = 2$, but if \TK\ becomes of the order of $\Delta_{CEF}$, higher values for $N$ should be considered~\cite{Tsujii2005}.

\begin{figure}[t]
\centering
\includegraphics[width=1.0\columnwidth]{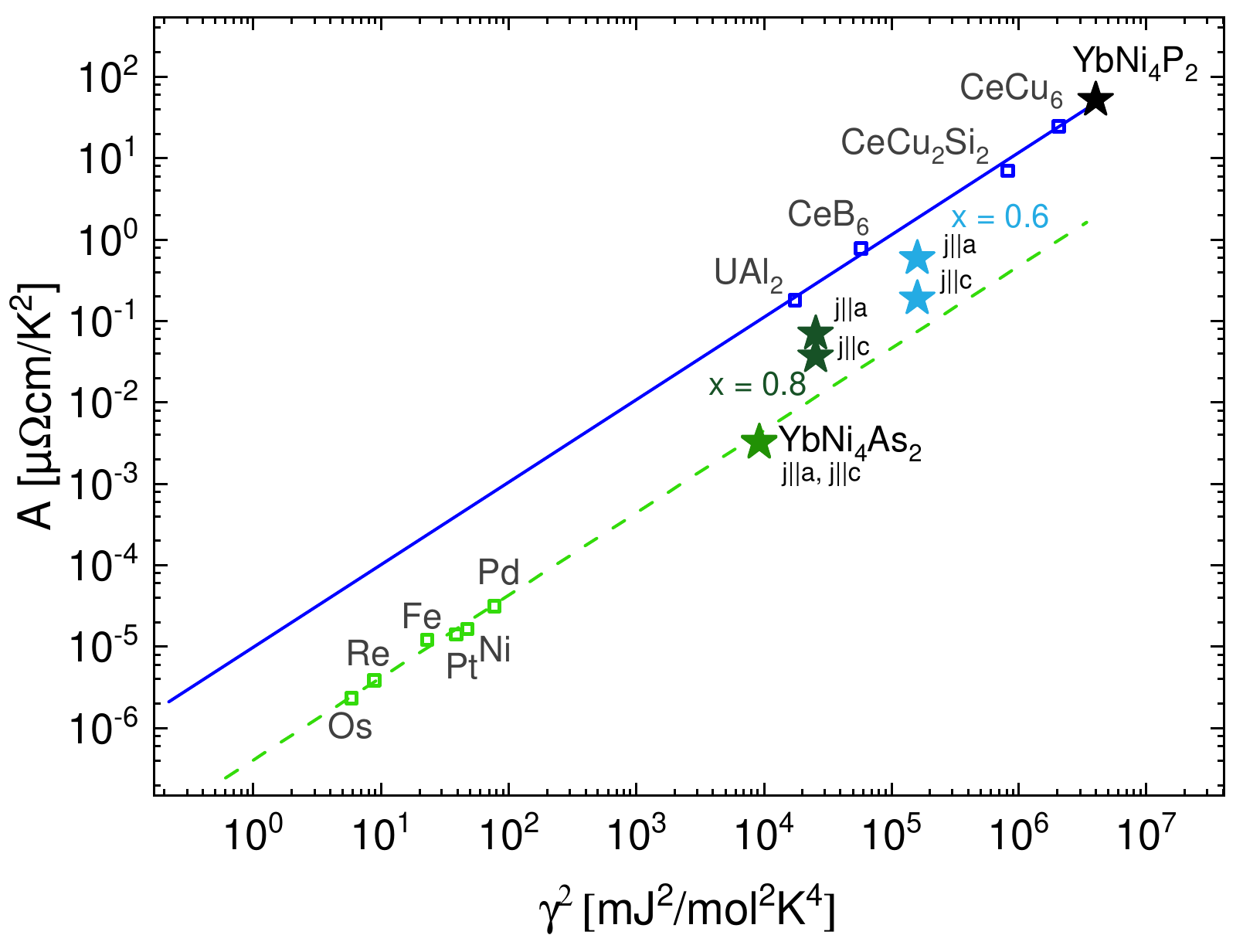}
\caption{Kadowaki-Woods ratio for some crystals of \YNPA\ (marked by stars) compared to other HF systems and transition metals, after~\cite{Jacko2009}.}
\label{KadowakiWoods}
\end{figure}

In Tab.~\ref{Sommerfeld} and Fig.~\ref{KadowakiWoods}, we compare the KW ratio of our crystals: While the pure phosphorous compound with $\gamma_0\approx 2000\,\rm mJ/\rm molK^2$, $A\approx 52\,\mu\Omega\rm cm/K^2$ and $A/\gamma_0^2=13\,\mu\Omega\rm cmK^2mol^2/J^2$ \cite{Krellner2011} is located in close vicinity to other HF compounds (solid line) with $N = 2$, \YNA\ is characterized by a much smaller ratio as typically found in transition metals without strong correlations (dashed line). Our compounds with $x = 0.8$ and $x = 0.6$ are situated inbetween the HF and the transition-metal lines. This can be attributed to the much higher \TK\ observed in these compounds compared to the pure \YNP . 
%

\begin{figure}
\includegraphics[width=1.0\columnwidth]{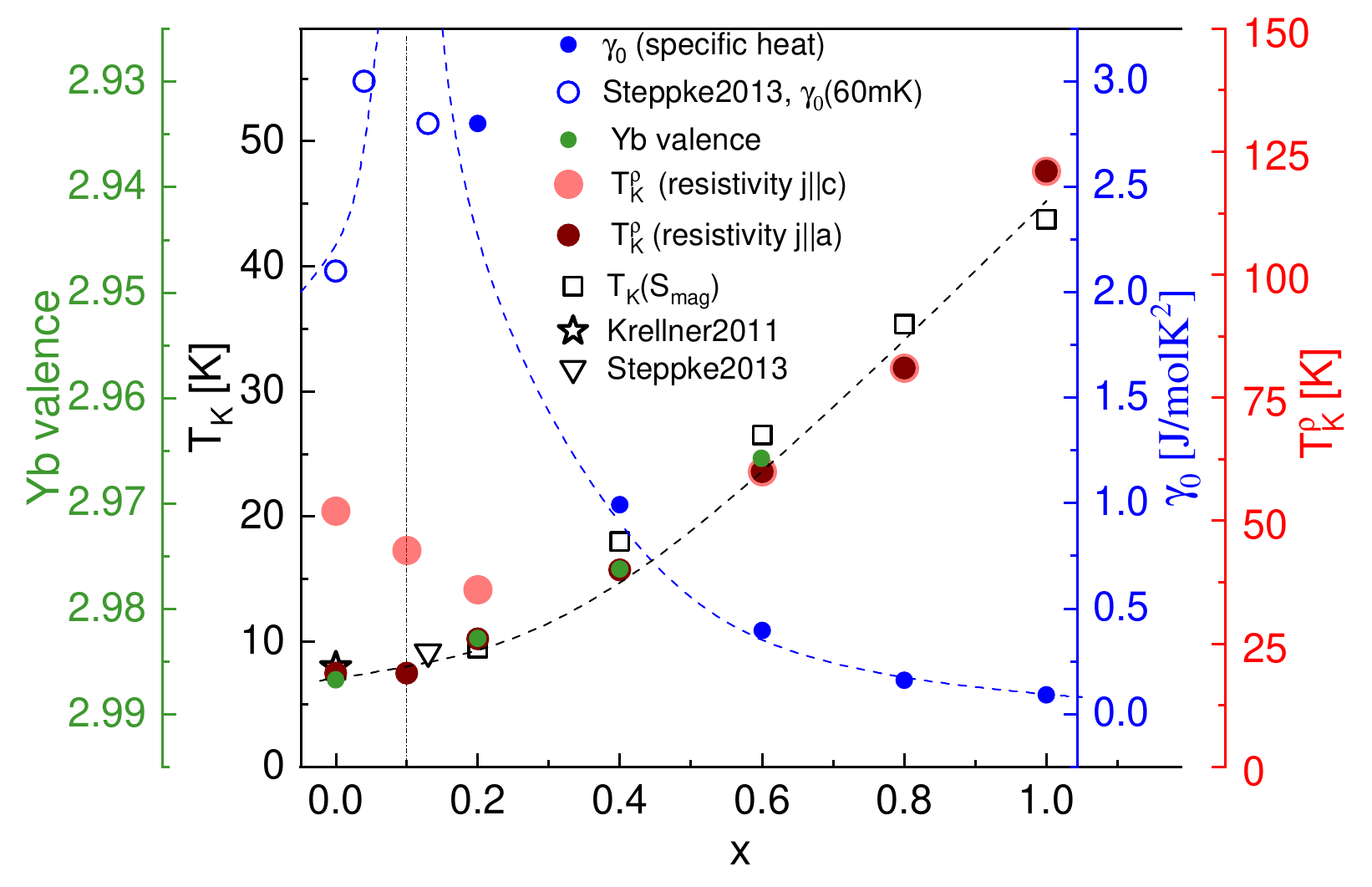}
\caption{Comparison of the characteristic energy scales and parameters in \YNPA: the Kondo temperature \TK\ (black symbols) determined from the entropy (Fig.~\ref{HCYNPAs}), the Sommerfeld coefficient $\gamma_0$ (blue points),  the temperatures at which the resistivity shows a maximum $T^{\rho}_{\rm K}$ (red points, determined in Fig.~\ref{RhoARhoC_1}) and the Yb valence at low temperatures (green points) from RXES. Dashed lines are guides to the eyes. }
\label{Kondobild}
\end{figure}
%

The increase of \TK\ as well as the change of the characteristic parameters with increasing negative chemical pressure is illustrated in Fig.~\ref{Kondobild}: We plot the Kondo temperature \TK, determined from the entropy analysis, the Sommerfeld coefficient $\gamma_0$ and the temperatures at which the resistivity shows a maximum T$^{\rho}_{\rm K}$. While $C/T$ saturates below $5\,\rm K$ for $x=1.0$ and $0.8$, below $\approx 0.35\,\rm K$ for $x=0.6$ and below $0.06\,\rm K$ for $x=0.4$, this is not the case for $x = 0.13$ and $x = 0.2$. We do then plot for these concentrations the values measured at the lowest temperature of 35\,mK. We find that the Kondo temperature \TK\ determined from the entropy scales with the maximum in the resistivity $T^{\rho}_{\rm K}$ for $j\parallel a$ for all As concentrations while the behaviour is different for maxima in the resistivity for $j\parallel c$ up to $x=0.2$. \TK\ increases up to $\approx 44\,\rm K$ for $x = 1.0$. The increase of \TK\ is accompanied by a strong decrease of $\gamma_0$. In the sample with $x = 0.6$, Fermi liquid behaviour sets in below 3\,K in the electrical resistivity while the heat capacity further increases below this temperature. The same trend was observed in samples with $x=0.8$ and $1.0$. The temperature below which FL behaviour is found is always higher in the resistivity than in the heat capacity (see Tab.~\ref{Resistivitycoeff}).
%

%
\section{Summary}
We used the Czochralski method to grow nine single crystals of the series \YNPA\ with $0\leq x \leq 1.0$ from a self flux. Energy dispersive X-ray spectroscopy measurements indicated that the phosphorous to arsenic ratio of the respective melt is maintained in the crystals and the As distribution is homogeneous throughout the whole sample. Rietveld refinement of the PXRD patterns showed that with increasing As content the $c$-lattice parameter increases stronger than the $a$-lattice parameter. The unit cell volume of \YNA\ is 10\,\% larger than that of \YNP.

We performed measurements of the magnetization, electrical resistivity, thermopower, heat capacity and resonant X-ray emission spectroscopy to characterize all crystals and deduce the $x - T$ phase diagram of \YNPA. All data consistently show 
that with increasing $x$ the Kondo temperature \TK\ increases, whereas the CEF scheme is less affected by the chemical pressure, as expected. A continuous change in Yb valence from nearly trivalent at low $x$ to a slightly lower value for those with higher $x$ was measured. Nevertheless, all high-$T$ susceptibility measurements can be well fitted with effective moments close to 4.54\,\muB\, which corresponds to the $J = 7/2$ multiplet of the free Yb$^{3+}$ ion. So, the system is still in the Kondo regime and away from being in the mixed-valence state. 

At low temperature we observed that the strong anisotropy found in the magnetic susceptibility and electrical resistivity in samples with low As content disappears for samples with higher As concentrations. Also FM correlations become weaker. Consequently, at low temperatures, the system shows a crossover from pronounced NFL behaviour for $x \leq 0.2$, due to the presence of a FM QCP, to a FL behavior for $x > 0.2$ with weak correlations and much lower Kadowaki-Woods ratios. Interestingly, heat capacity data at very low temperatures show that $C/T$ strongly increases towards lower $T$ with a similar power law for $x = 0.13$ and $x = 0.2$. This suggests that in \YNPA\ a quantum critical region rather than a quantum critical point might exist. Within this region, quantum Griffiths phase effects - as observed in other FM QC systems like CePd$_{1-x}$Rh$_{x}$~\cite{Westerkamp2009} or Ce(Pd$_{1-x}$Ni$_{x}$)$_{2}$P$_{2}$~\cite{Lai2018} - can be excluded because the power-law divergence in specific heat is insensitive to the amount of disorder $x$. In fact, it is almost the same for all $x \leq 0.2$. It might be possible that in quasi-1d Kondo lattice ferromagnets ~\cite{Komijani2018} disorder has a weaker effect than in 2d or 3d systems~\cite{Vojta2010}. 
%
\begin{acknowledgments}
KK and CK thank C. Geibel, W. A\ss mus and F. Ritter 
for valuable discussions, and K.-D. Luther for technical support. The authors grateful acknowledge support by the Deutsche Forschungsgemeinschaft (DFG) through grants KR 3831/4-1 and BR 4110/1-1.
We acknowledge the European Synchrotron Radiation Facility (ESRF) for provision of synchrotron radiation facilities under proposal number IH-HC-3460.\\
\end{acknowledgments}
%
\cleardoublepage
\newpage
\noindent
\textbf{\Large Supplemental Material}
\renewcommand{\thefigure}{S\arabic{figure}}
\setcounter{figure}{0}

\subsection{Crystal growth}
\begin{figure}[bhtp]
\centering
\includegraphics[width=1\columnwidth]{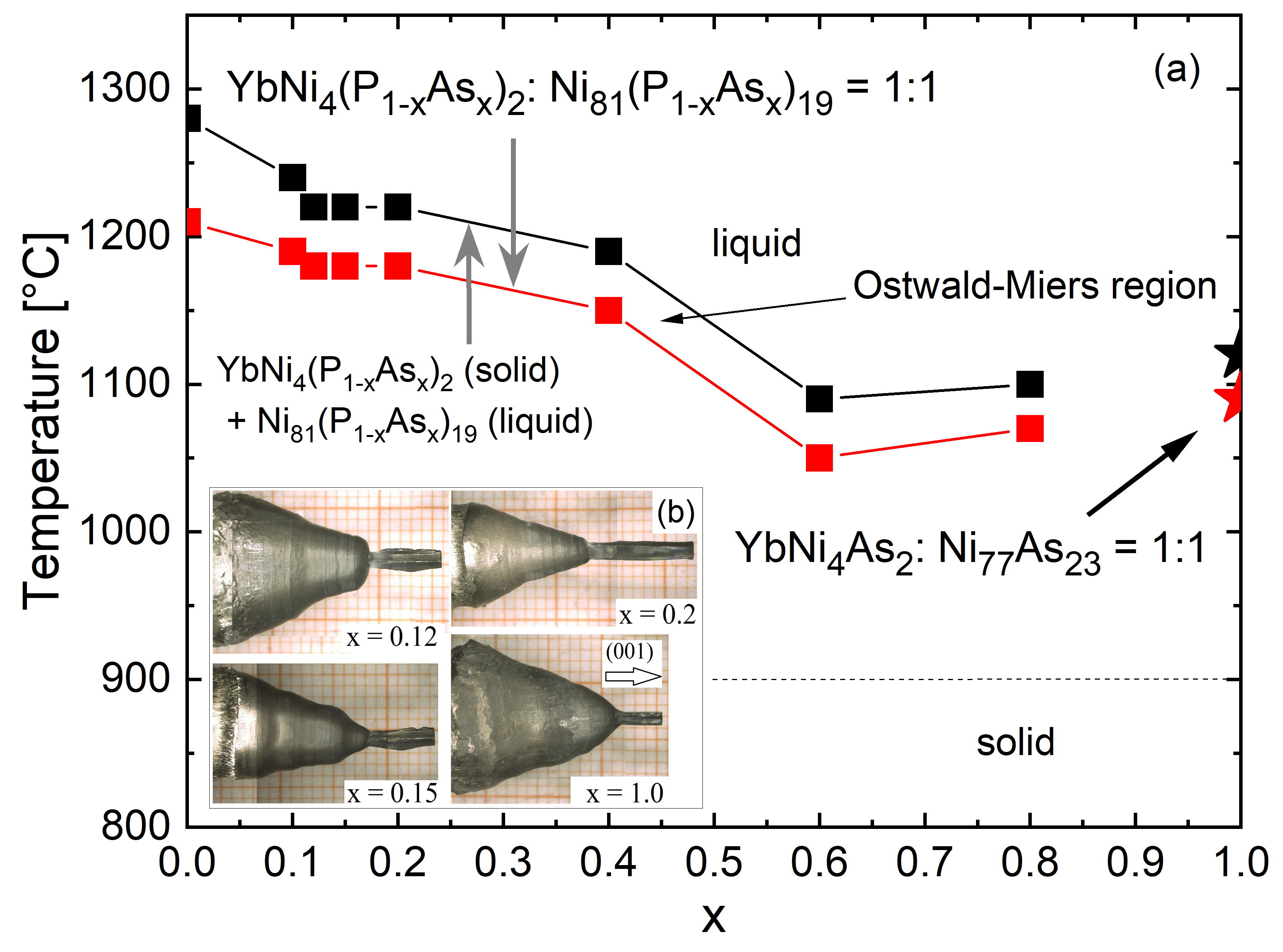}
\caption{(a) Melt composition - temperature phase diagram for the growth of As substituted YbNi$_4$P$_2$ single crystals. Black/red symbols mark the temperatures where the first signs of melting/solidification are observable during heating/cooling of the precursor of different As content ${\it x}$. Squares indicate the usage of the Ni$_{81}$P$_{19}$ eutectic and stars the usage of the Ni$_{77}$As$_{23}$ eutectic as flux. Gray arrows mark the direction of temperature change to obtain the data points. (b) Resulting YbNi$_4$(P$_{\rm 1-x}$As$_{\rm x}$)$_2$ single crystals.}
\label{solidliquid}\label{singlecrystal}
\end{figure}

YbNi$_4$(P$_{\rm 1-x}$As$_{\rm x}$)$_2$ single crystals were grown using the Czochralski method from a levitating melt according to the procedure described in \cite{Kliemt2016}. For each melt composition, liquidus and solidus temperatures according to Fig.~\ref{solidliquid}(a) were determined pyrometrically for all experiments by directly observing the beginning of the melting or solidification through the front window of the growth apparatus. The samples Fig.~\ref{solidliquid}(b) are nearly inclusion free in the first $\approx 8\,\rm mm$ that were pulled out of the melt.

\subsection{Characterization}
\begin{figure}[bhtp]
\centering
\includegraphics[width=1.0\columnwidth]{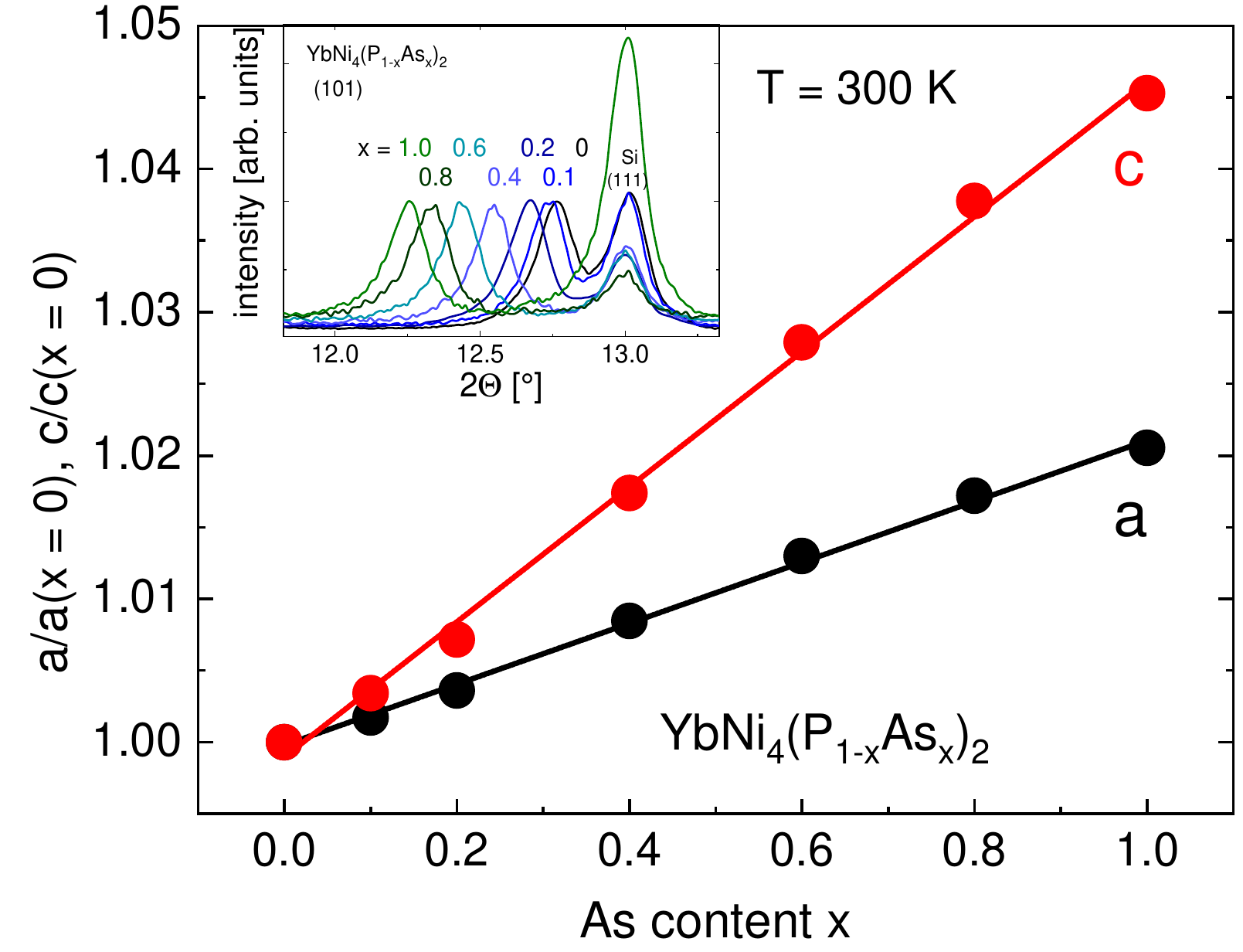}
\caption{Normalized lattice parameters $a$ and $c$ of YbNi$_4$(P$_{1-{\it x}}$As$_{\it x}$)$_2$ from Tab.~\ref{Gitterkonstanten} determined by Rietveld refinement. The size of each point exceeds the statistical error. {\it Inset:} X-ray powder diffraction data of the (1 0 1) reflex measured at 300 K with various substitution levels. }
\label{lattice}\label{peakshift}
\end{figure}

The lattice parameters of the samples in the substitution serious displayed in Tab.~\ref{Gitterkonstanten} follow Vegards law and are depicted in Fig.~\ref{lattice}. The inset shows the shift of the (1 0 1) reflex for increasing As substitution level indicating an enlargement of the unit cell. Silicon was used as standard and the position of the Si (1 1 1) is shown for comparison.

\cleardoublepage

\subsection{Inverse magnetic susceptibility}

\begin{figure}[htpb]
\includegraphics[width=1.0\columnwidth]{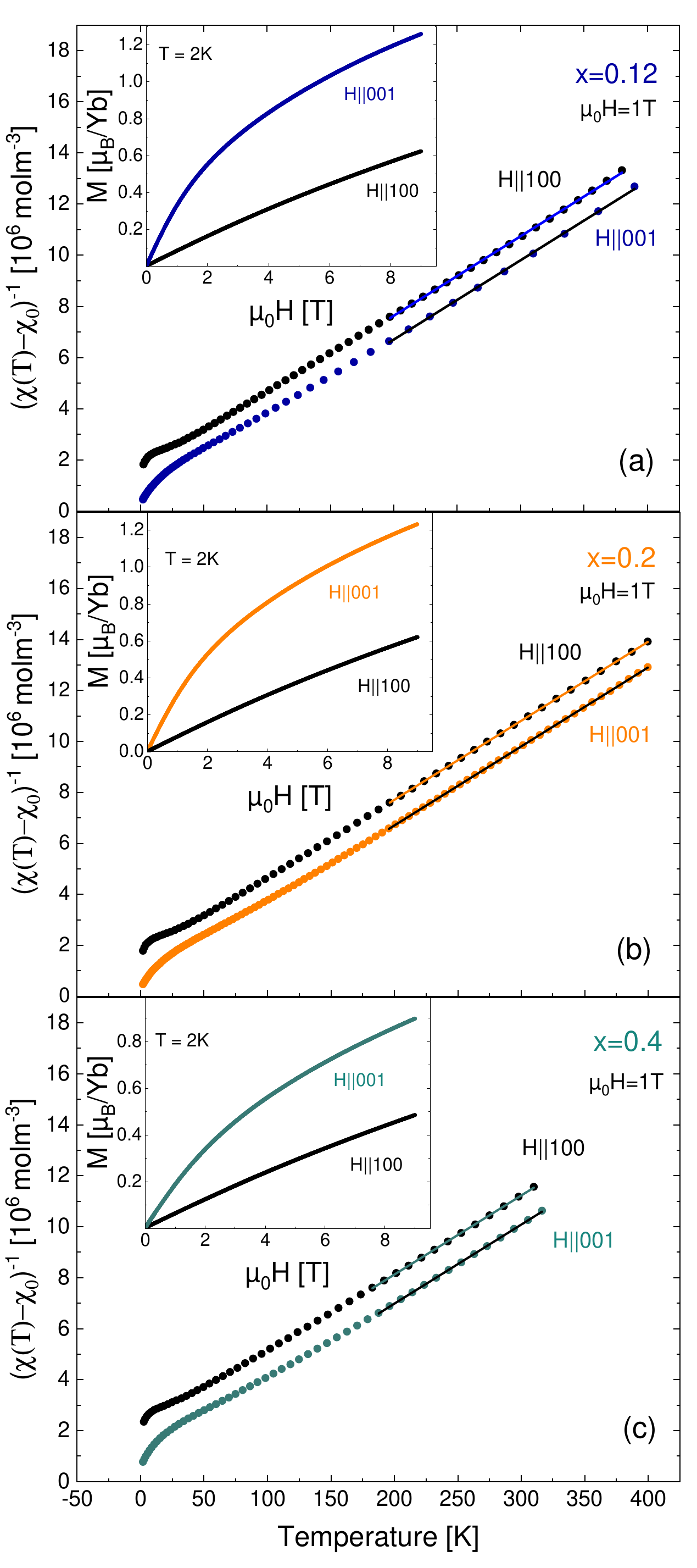}
\caption{The inverse magnetic susceptibility $(\chi(T)-\chi_0)^{-1}$ with $\mu_0H = 1\,T$ for  (a) $x=0.12$, (b) $x=0.2$ and (c) $x=0.4$ presents paramagnetic Curie-Weiss behaviour above $200\,\rm K$. In the insets, the field dependence of the moment per Yb ion at $T=2\,\rm K$ for the two field directions $H\parallel 001$ and $H\parallel 100$ is shown. 
}
\label{MvT_YBNi4PAs2_1}\label{YNPAs_mueff}
\end{figure}

\begin{figure}[htpb]
\includegraphics[width=0.9\columnwidth]{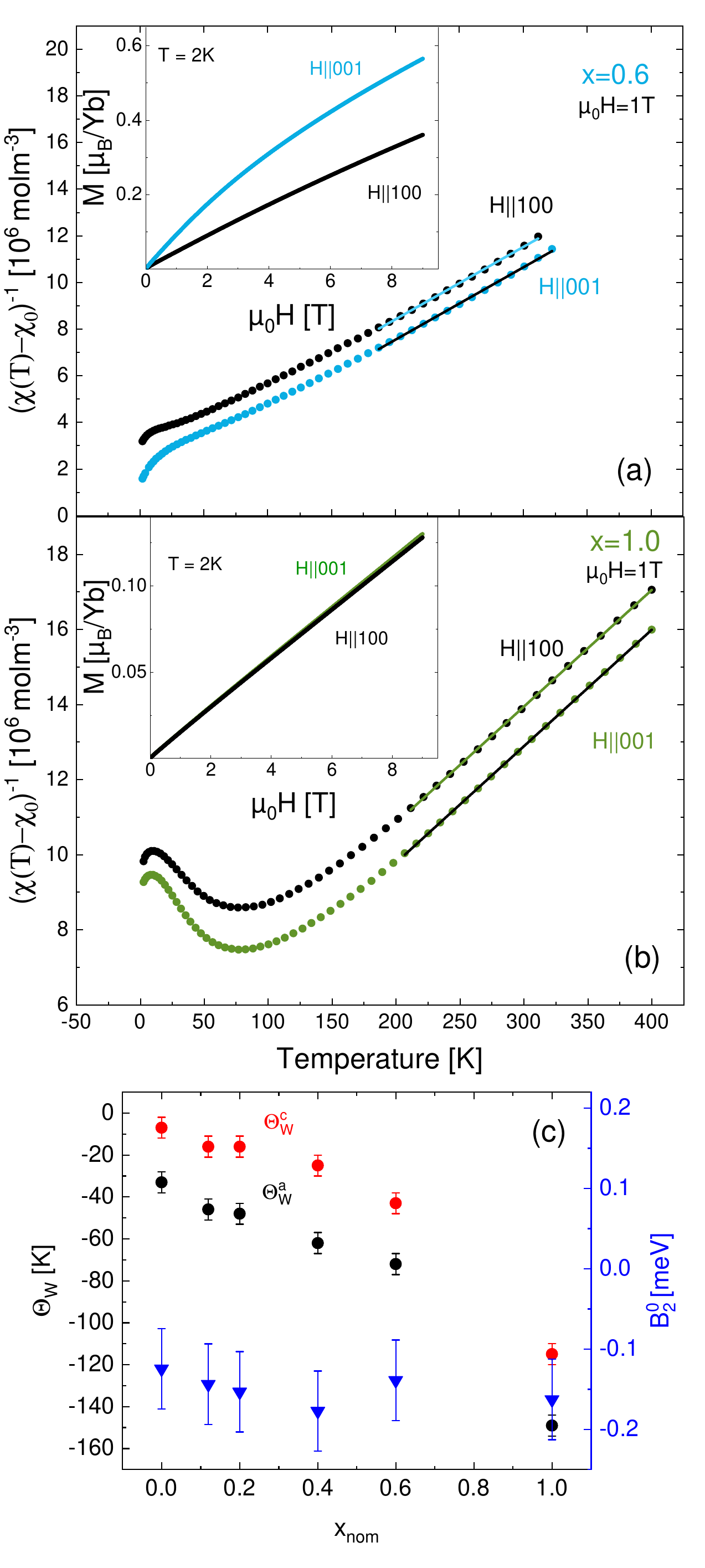}
\caption{YbNi$_4$(P$_{1-{\it x}}$As$_{\it x}$)$_2$ temperature dependence of the inverse magnetic susceptibility $(\chi(T)-\chi_0)^{-1}$ for $\mu_0H = 1\,T$ (a) $x=0.6$ and (b) $x=1$ presenting paramagnetic Curie-Weiss behaviour above $200\,\rm K$. In the insets, the field dependence of the moment per Yb ion at $T=2\,\rm K$ for the two field directions $H\parallel 001$ and $H\parallel 100$ is shown. (c) YbNi$_4$(P$_{1-{\it x}}$As$_{\it x}$)$_2$ Weiss temperatures (circles) and the first CEF parameter $B^0_2$ (triangles) versus nominal As concentration $x$. This plot shows the data from Tab.~\ref{TabelleThetaW}.  
See Eqn.~(\ref{B02}) and \cite{Klingner2011} for definition of $B^0_2$. }
\label{MvT_YBNi4PAs2_1a}\label{YNPAs_mueffa}\label{WeissT}
\end{figure}

Figs.~\ref{MvT_YBNi4PAs2_1}(a-c) and \ref{MvT_YBNi4PAs2_1a}(a,b) show the inverse magnetic susceptibility using the data from Fig.~\ref{MvT_YBNi4PAs2}. 
For the samples with high As content $x\geq0.4$, a small Pauli susceptibility contribution $\chi_0$ was subtracted such that the slope of the high-temperature Curie-Weiss fit matches that of the free Yb$^{3+}$ ion. 

\subsection{Resistivity}

\begin{figure}[htpb]
\centering
\includegraphics[width=0.97\columnwidth]{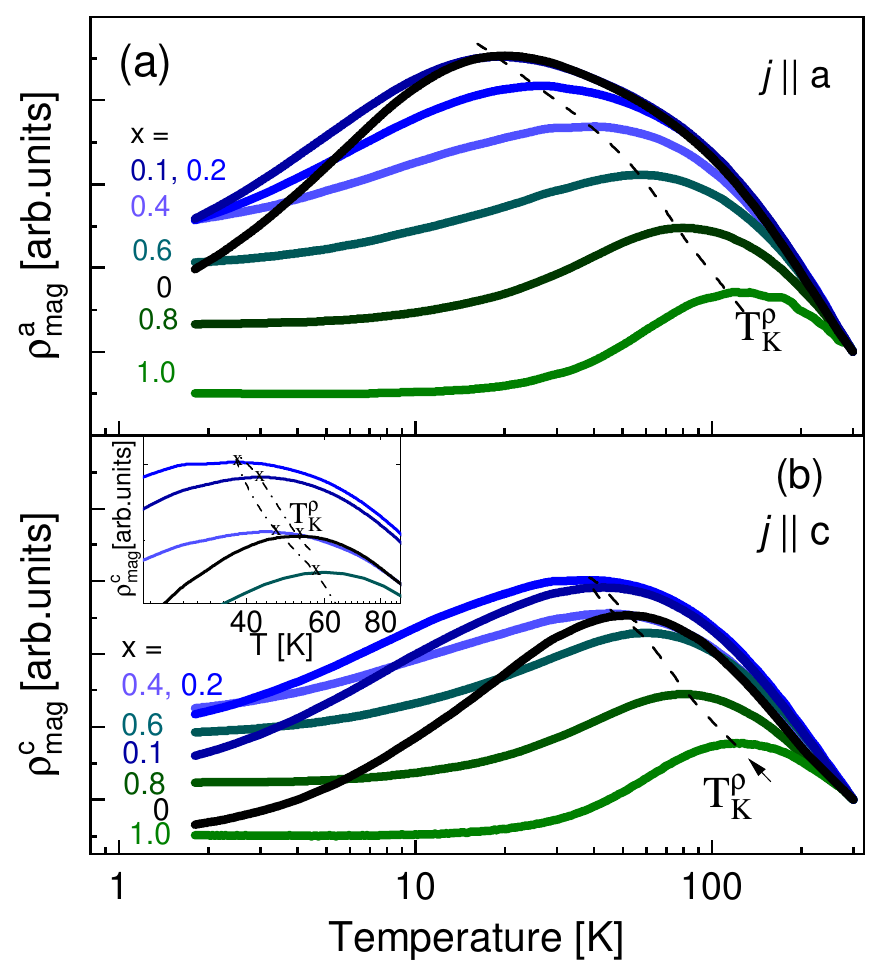}
\caption{Figures (a) and (b) show $\rho_{\rm mag}(T)$ obtained by subtraction of the LuNi$_4$P$_2$ reference data from the measured resistivity data, Fig.~\ref{RhoARhoC}, according to Eqn.~(1). {\it Inset:} Loop-like shift of the maximum for $j\parallel 001$ which might be an artefact due to the method of data evaluation using the  nonmagnetic background.}
\label{RhoARhoC_1}
\end{figure}

$\rho_{\rm mag}(T)$ is depicted in Fig.~\ref{RhoARhoC_1}.
In the data obtained on YbNi$_4$(P$_{1-x}$As$_x$)$_2$, $x\leq 0.4$, besides the maximum at $T^{\rho}_{\rm K}$, a high-temperature shoulder at $\approx 100\,\rm K$ is visible which does not change its position and merges completely with the maximum for higher substitution levels. Note that for instance for YbRh$_2$Si$_2$ under pressure the occurrence of a distinct high-temperature maximum was observed and attributed to inelastic Kondo scattering on excited CEF levels as described in \cite{Dionicio2005}.

\subsection{Heat capacity}
\begin{figure}[htpb]
\centering
\includegraphics[width=1.0\columnwidth]{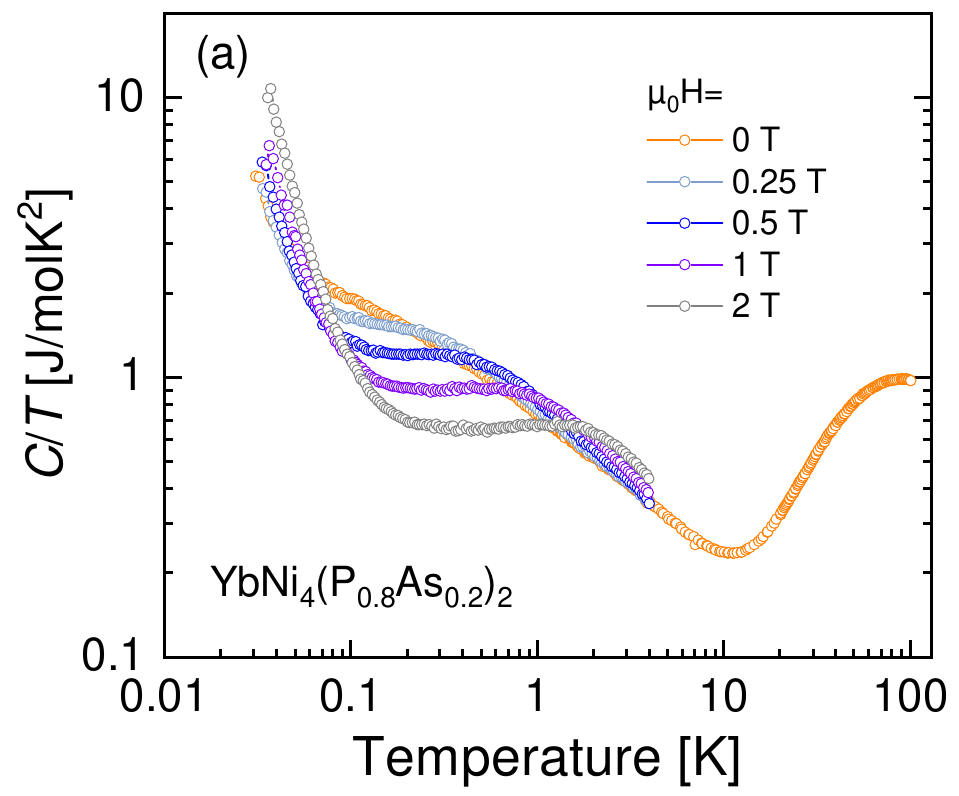}
\includegraphics[width=1.0\columnwidth]{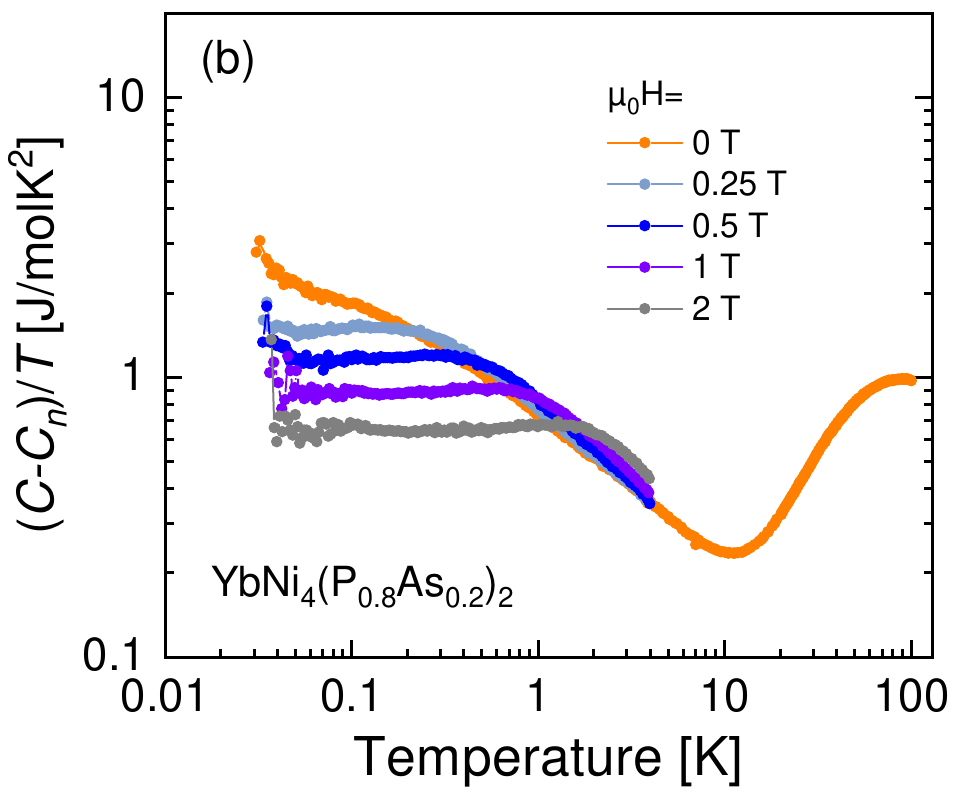}
\caption{Heat capacity divided by temperature of a sample with $x=0.2$. (a) $C/T$ versus $T$ in different magnetic fields. (b) $(C-C_n)/T$ versus $T$.  }
\label{HC20proz_nucl}
\end{figure}

\begin{figure}[htpb]
\centering
\includegraphics[width=1.0\columnwidth]{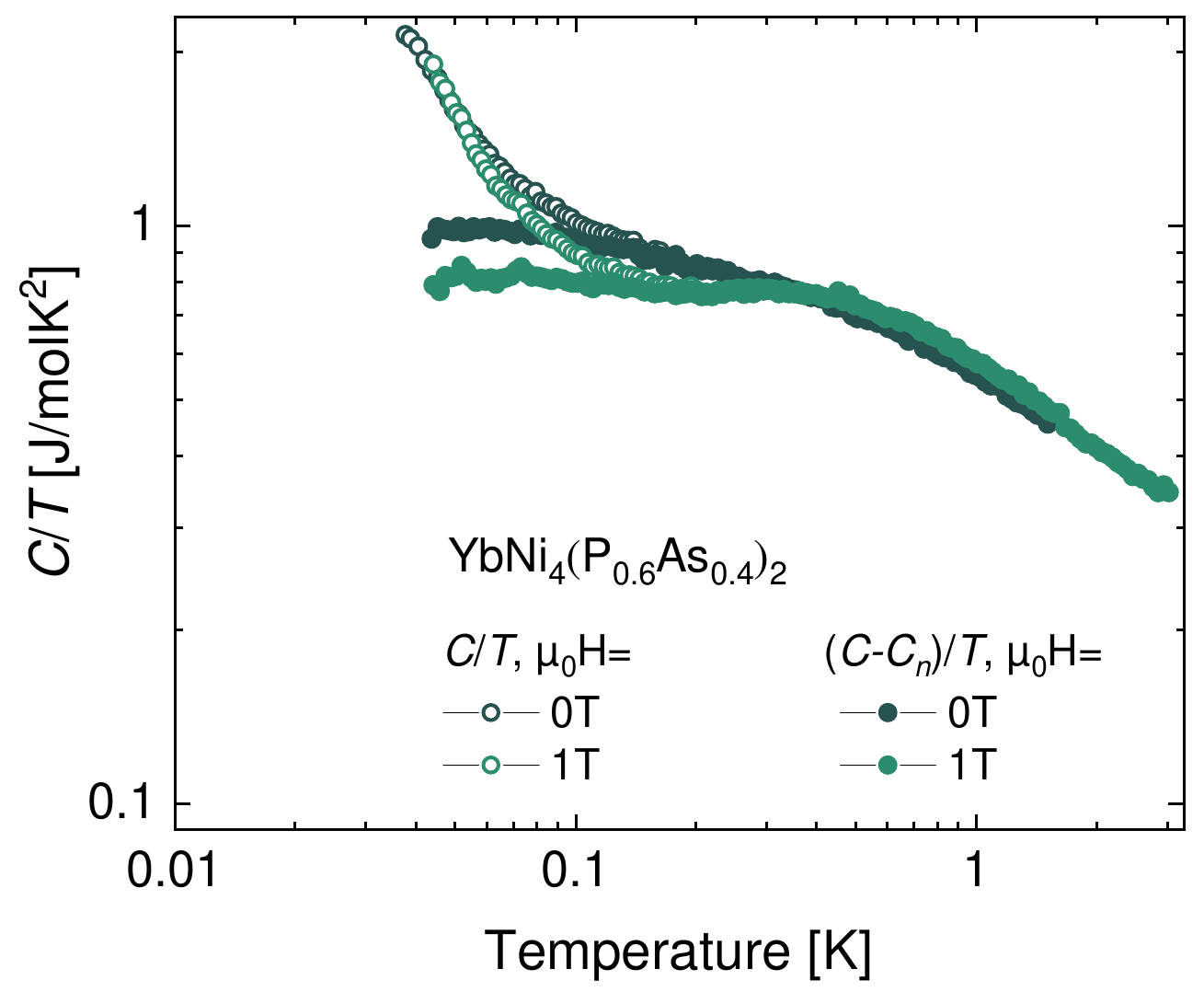}
\caption{$C/T$ versus $T$ of a sample with $x=0.4$ in zero magnetic field and in a field of $\mu_0H=1\,\rm T$ (open symbols) and  $(C-C_n)/T$ versus $T$ (closed symbols). }
\label{HC40proz_nucl}
\end{figure}

Figs.~\ref{HC20proz_nucl} and \ref{HC40proz_nucl} show the heat capacity divided by temperature for samples with $x=0.2$ and $x=0.4$ respectively.
For both samples $(C-C_n)/T$ versus $T$ is displayed with the nuclear contribution $C_n$ subtracted according to the procedure described in \cite{Steppke2013}.

\bibliography{YbNi4PAs2.bib}
\bibliographystyle{apsrev4-2}
\end{document}